\documentstyle[psfig]{mn}

\title[Dwarf Nova Oscillations and Quasi-Periodic Oscillations in Cataclysmic Variables: III.]
{Dwarf Nova Oscillations and Quasi-Periodic Oscillations in Cataclysmic Variables: III. A New
Kind of Dwarf Nova Oscillation, and Further Examples of the Similarities to X-Ray Binaries.}
\author[Brian Warner, Patrick A. Woudt and Magaretha L. Pretorius]
       {Brian Warner\thanks{email: warner@physci.uct.ac.za}, 
        Patrick A. Woudt\thanks{email: pwoudt@circinus.ast.uct.ac.za} and
        Magaretha L. Pretorius\thanks{email: retha@mensa.ast.uct.ac.za} \\
        Department of Astronomy, University of Cape Town, Private Bag,
        Rondebosch 7700, South Africa}
\date{2003 April 14}

\begin{document}

\maketitle

\begin{abstract}
   We present measurements of the periods of Dwarf Nova Oscillations 
(DNOs) and Quasi-Periodic Oscillations (QPOs) in cataclysmic variable 
stars (CVs), many culled from published literature but also others newly 
observed (in VZ Pyx, CR Boo, OY Car, Z Cha, AQ Eri, TU Men, HX Peg, CN Ori, V893 Sco, WX Hyi and EC2117-54). 
These provide data for 26 systems. We show that in 
general $P_{QPO}$ $\sim$ 15 $P_{DNO}$ and that the correlation for CVs extends by three 
orders of magnitude lower in frequency the similar relationship found for 
X-Ray binaries. In addition, we have found that there is a second type of DNO, 
previously overlooked, which have periods $\sim$ 4 times those of the 
regular DNOs (As well as those mined from publications, we have observed them
in VW Hyi, OY Car, AQ Eri, V803 Cen, CR Boo, VZ Pyx, HX Peg and EC2117-54). 
Often both types of DNO coexist. Unlike the standard DNOs, 
the periods of the new type, which we refer to as longer period DNOs 
(lpDNOs), are relatively insensitive to accretion luminosity and can even 
appear in quiescence of dwarf novae. We interpret them as magnetically 
channelled accretion onto the differentially rotating main body of the white 
dwarf primary, rather than onto a rapidly slipping equatorial belt as in the 
case of the standard DNOs. This is supported by published measurements of 
$v \sin{i}$ for some of the primaries.        
     Some similarities of the DNOs, lpDNOs and QPOs in CVs to the three 
types of QPO in X-Ray binaries (burst pulsations, high and low frequency 
QPOs) are noted.

\end{abstract}


\section{Introduction}
   Dwarf nova oscillations (DNOs) and quasi-periodic oscillations (QPOs) in 
cataclysmic variables (CVs) have been observed for thirty years -- the DNOs 
were discovered in 1972 (Warner \& Robinson 1972), the QPOs were first 
recognised in 1977 (Patterson, Robinson \& Nather 1977) and were found to 
be clearly discernable in some earlier light curves.

    The accumulated observations of DNOs and QPOs have been listed in 
Warner (1995a) and discussed there and in the first two papers of this series 
(Woudt \& Warner 2002: Paper I; Warner \& Woudt 2002a: Paper II; see also 
Warner \& Woudt 2002b, 2003). The observations of VW Hyi presented and 
analysed in Paper I, and the discussion given in Paper II, led to the 
realization that in that star the period ratio $R$ = $P_{QPO}$/$P_{DNO}$ remains 
approximately constant at a value $\sim$ 15 during the rapid increases in these 
periods seen near the end of VW Hyi outbursts. We pointed out that not only 
is $R$ similar to that found for the high and low frequency QPOs observed in 
X-Ray binaries (Psaltis, Belloni \& van der Klis 1999), the VW Hyi 
relationship in fact appears to be an extension of the X-Ray binary 
correlation to frequencies two orders of magnitude lower.

     Subsequently, Mauche (2002) discovered that DNOs and QPOs observed 
in the EUV and soft X-Ray regions during an outburst of the dwarf nova SS 
Cyg have $R$ $\sim$ 12, and the DNO periods ($\sim$ 8 s) place them between VW Hyi 
and the X-Ray binaries in the two-QPO correlation diagram. This greatly 
strengthened the evidence for the existence of a general relation between high 
and low frequency QPOs in all of these interacting binaries, irrespective of 
whether the primary is a black hole, neutron star or white dwarf. In CVs the 
DNOs are the analogues of the high frequency X-Ray binary QPOs, and the 
CV's QPOs are the analogues of the low frequency X-Ray QPOs.

    In this paper we will be largely interested in presenting new observations 
of DNOs and QPOs in CVs, and in reinterpreting previously published 
observations. A complication that arises is that both DNOs and QPOs can 
show frequency doubling, and these can happen independently of each other 
(Warner \& Woudt, in preparation). This may change the measured value of 
$R$ by a factor of 2 in some cases. QPOs are far less coherent and persistent
than DNOs. It is commonly observed that a train of 5 to 10 moderately coherent
QPO cycles is clearly visible, before changing in phase or disappearing
altogether.

     We first indicate some of the theoretical work that appears relevant to 
understanding the nature of the rapid oscillations in interacting binaries.
   
     That DNOs are signatures of magnetically controlled accretion onto the 
equatorial belt of the white dwarf primaries of CVs was proposed by 
Paczynski (1978), made more quantitative by Warner (1995b) and 
elaborated in Paper II as the Low Inertia Magnetic Accretor (LIMA) model. 
The last also develops the idea that QPOs may be caused by prograde 
travelling waves near the inner edge of the magnetically truncated accretion 
discs. A simple excitation mechanism, involving magnetic reconnection, 
leads to the expectation that $R \sim {\Omega_{k}}/{\Delta \Omega} \sim
(1 - \omega_{s})^{-1}$, where $\Omega_k$ is the 
Keplerian frequency in the disc near the corotation radius $r_{co}$, 
$\Delta \Omega = \Omega_k - \Omega_{\rm max}$, where $\Omega_{\rm max}$
is the maximum angular frequency in the transition 
zone between $r_{co}$ and the inner edge of the disc, and $\omega_s$ is 
the  `fastness parameter' defined by Ghosh \& Lamb (1979).

    A somewhat different physical interpretation is that QPOs are 
modulations in the rate of mass transfer onto the primary. This concept is 
used in QPOs of the luminosities of young stars, where the QPOs are 
envisaged as magnetospheric radial oscillations, caused by field winding and 
reconnecting, as in the above proposed model for CVs. The radial 
oscillations produce quasi-periodic modulation of the mass transfer rate, and 
a typical model has $R \sim 100/{2 \pi}$ (Goodson \& Winglee 1999).

    Returning to interacting binaries, in a study of field winding, inflation and 
reconnection in the region connecting the inner disc to the primary, 
Uzdensky (2002) concludes that quasi-periodic oscillations should occur. 
Such QPOs could cause mass transfer modulation and/or travelling waves. 
Titarchuk \& Wood (2002) interpret the whole range of the two-QPO 
correlation as evidence for magnetoacoustic oscillations in the transition 
zone, driven by the adjustment to sub-Keplerian flow.

    The next two sections of this paper are devoted to accumulating values of 
$P_{DNO}$ and $P_{QPO}$ and the resulting estimates of $R$: In Section 2 we examine 
published papers in order to extract values of $R$; in Section 3 we describe 
new photometric observations of CVs that lead to further determinations of 
$R$. In Section 4 we draw attention to a new type of DNO that is occasionally 
seen, which could be confused with the standard DNOs, and in Section 5 we 
compare the DNOs and QPOs in CVs with the two-QPO correlation and 
other properties of X-Ray binaries. Section 6 contains a final discussion and 
assessment.

\section{Published observations of DNOs and QPOs}

    The values of $R$, their possible variation for any given star, and their range 
from star to star, have not previously been given much attention, but in the 
light of the newly discovered possible connection between CV and X-Ray 
QPOs these ratios are of great interest. In this Section we examine the 
published studies of CVs with this in mind. Ideally, we would prefer light 
curves where both DNOs and QPOs are present simultaneously. However, 
largely through the lower interest in QPOs in the past, DNO periods are 
often the only ones quoted. We can expand the list of useful results by 
including observations of `double' DNOs, making use of the conclusion 
(Papers I and II) that in at least some CVs (VW Hyi, WZ Sge) the QPO frequency is 
the difference (i.e. beat) frequency of the two DNOs. The true DNO is the 
shorter of the two periods; the longer period is the result of reprocessing of 
the DNO by the travelling wave that is thought to produce the QPO (Paper 
II). By using the beat frequency we in effect generate a `pseudo' QPO; this 
may not be visible directly in the light curve because the travelling wave 
may happen not to intercept our line of sight, but its presence is indirectly 
observed by the shielding and reprocessing effects it has on the disc, which 
generate the double DNO.

\subsection{RU Peg}

    The CV in which QPOs were first noticed was the dwarf nova RU Peg. 
Patterson et al.~(1977) found $\sim$ 50 s oscillations clearly discernable in the 
light curve which, because of their low coherence, were transformed into a 
barely significant broad and noisy band of power in the Fourier transform 
(FT). The QPOs were present during all five nights of the November 1975 
outburst. In addition, on four of the nights more stable oscillations, 
characteristic of DNOs, were seen in the FTs at periods near 11.6 s but with 
unusually small amplitudes $\sim$ 0.6 mmag (millimagnitudes).

   The value of $R$ for RU Peg, using the above numbers, is $\sim$ 4.5. This is so 
far removed from the value $\sim$ 15 that we see in VW Hyi and the X-Ray 
binaries that it would seem that the QPO `type star' itself is highly 
anomalous. However, the following reasoning leads to a different 
conclusion. The QPOs at $\sim$ 50 s are clearly present; taking these as the 
starting point we could look for DNOs at a period $\sim$ 50/15 $\sim$ 3.3 s. The 
integration time used by Patterson et al.~was 4 s, so the `predicted' DNOs are 
above their Nyquist frequency. However, if a $\sim$ 3.3 s signal were actually 
present in the RU Peg light curve, it could show up at the beat frequency 
between $\sim$ 3.3 s and 4 s. This is of the order of the period of the DNOs 
actually observed in RU Peg. The relatively long integration time resulted in 
a reduction of amplitude by a factor of $\sin{x}/x$, where $x = \pi \delta{t}/P_{DNO}$ 
and $\delta{t}$ is the integration time. Thus the true DNO amplitude in RU Peg would be a 
factor 4.7 larger, or 2.8 mmag, which is more in line with what is commonly 
observed in dwarf novae near maximum of outburst.

     The value of the ratio $R$ for RU Peg, deduced from the posited DNO 
period (at 2.97 s -- the alias of 11.6 s) and the QPO period, is 16.8.

     A few years ago DNOs at $\sim$ 3 s would have seemed unlikely or 
impossible. But the observation of $\sim$ 3 s DNOs in the EUV and X-Ray 
regions of SS Cyg during outburst (Mauche 1998; van Teeseling 1997) 
should remove that pre\-judice. In SS Cyg the DNOs had undergone a sudden 
frequency doubling, and it is possible that the same happened in RU Peg. 
We note that DNOs are associated with the Keplerian period close to the 
primary, and that it requires large masses to produce these short periods. The 
mass of the white dwarf in SS Cyg is indeed quite large -- 1.20 M$_{\odot}$
according to Robinson, Zhang \& Stover (1986) -- and the mass of the 
primary in RU Peg is 1.38 M$_{\odot}$ (Friend et al.~1990).

It is possible that the 11.6 s oscillation is an example of the lpDNOs
discussed in Section 4 of this paper. But the implication would be that any true
DNO would appear at $\sim$ 3 s, which does not change the implied value of $R$.

\subsection{V3885 Sgr}

    Hesser, Lasker \& Osmer (1974) observed 29.08 s and 30.15~s DNOs in 
the nova-like V3885 Sgr. No QPOs were commented on, so we use the 
estimate $P_{QPO}$ = 819 s from the beat period, to deduce $R$ = 28.2.

\subsection{WZ Sge}

    The 27.87 s and 28.95 s DNOs and the 742 s QPO seen in the light curve 
of WZ Sge (they are intermittently present outside of outburst: these are 
described in Paper II), lead to $R$ = 26.6.

\subsection{SS Cyg}

   The EUVE observations of SS Cyg in outburst by Mauche (2002) give a 
DNO period of 7.7 s and a QPO period of 83~s; his soft X-Ray (Chandra) 
observations give DNOs at 9.1~s and a QPO at 111 s. From his analysis, 
Mauche finds $R$ = 10.4 and $R$ = 11.4 respectively.

\subsection{V436 Cen}

   The outburst light curve in which DNOs were detected with an average 
period of 19.6 s (Warner 1975) has been further analysed and we find strong 
QPOs with $P_{QPO}$ $\sim$ 475 s.  This gives $R$ $\sim$ 24. Also, in one section of the light 
curve obtained on 11 June 1975 there is a double DNO with periods 19.45 s 
and 20.20 s (and a QPO at $\sim$ 500 s was visible at this time). Using the double 
DNO we get $R$ = 26.

\subsection{V2051 Oph}

    DNOs at 56.12 s, 28.06 s and 29.77 s during the late decline phase of a 
dwarf nova outburst were found in V2051 Oph by Steeghs et al.~(2001), and 
they observed the beat at 486 s. This gives $R$ = 17.3.

\subsection{UX UMa}

    Warner \& Nather (1972) and Nather \& Robinson (1974) found DNOs in 
the nova-like UX UMa, with periods near 29~s. As pointed out in Paper II, 
the O -- C diagram shown by Nather \& Robinson contains $\sim$ 650 s QPOs in 
it. These give $R \sim 22$.

\subsection{IX Vel}

    The bright nova-like IX Vel frequently has DNOs over the range 24.6 -- 29.1 s 
(Warner, O'Donoghue \& Allen 1985). Williams \& Hiltner (1984) 
made an autocorrelation analysis of extensive high speed photometry and 
found evidence for the presence of a persistent $\sim$ 500 s time scale. The latter 
is probably caused by QPOs, which give $R \sim 18$.

\subsection{TY PsA}

   Light curves of TY PsA during outburst have shown strong DNOs near 27 s 
(Warner, O'Donoghue \& Wargau 1989). During one run late on the 
descending branch of the outburst QPOs at 245 s were also present. 
Formally this gives $R \sim 9$, but it is possible that frequency doubling of the 
QPO had taken place.

\subsection{GK Per}
   
    GK Per is a particularly interesting object as it is an old nova that now 
shows dwarf nova outbursts. Furthermore, it has a clear 351 s primary spin 
period, $P_{spin}$, detected in X-Rays (Watson, King \& Osborne 1985), which 
shows up as $\sim$ 380 -- 440 s QPOs (which, for sake of consistency, we here 
call DNOs) after optical processing from some part of the accretion disc 
(Patterson 1981; Nogami, Kato \& Baba 2002). It also has longer period 
QPOs, around 5000 s, seen in X-Rays (Watson et al.~1985) and in the optical -- both 
in the continuum (Nogami et al.~2002) and in emission lines (Morales-Rueda, 
Still \& Roche 1996, 1999; hereafter MSR).

   Two very important conclusions are arrived at by MSR: first, that the 
optical modulation can be interpreted as variable radiation of the entire disc, 
from the `QPO radius' outwards, caused by the obscuration of the X-Ray 
illuminating source by a blob at the QPO radius in the inner disc; and 
second, that the QPOs arise from beating between more typical DNOs and 
the magnetic accretion curtain spinning with the white dwarf (as proposed 
by Watson et al.~1985). Both of these proposed structures are very similar to 
the LIMA model presented for DNOs and QPOs in Paper II, but with the 
difference that in GK Per the field of the white dwarf is high enough to 
prevent slippage of the accreting equatorial belt relative to the interior. That 
is, GK Per is a standard intermediate polar (IP), with the high inertia of the 
entire primary receiving the accretion torque. Nogami et al.~(2002) conclude 
that the unstable optical DNOs are probably the 351 s (constant period) 
rotating beam being reprocessed off the (varying period) QPO source, which 
in the LIMA model is a slow travelling wave moving progradely near the 
inner edge of the truncated disc.

   We feel it is appropriate, therefore, to include GK Per in the inventory of 
CVs possessing both DNOs and QPOs, and deduce that $R \sim 5000/380 = 13.2$.

   As pointed out in Paper II, in general it will be difficult to detect QPOs in 
the common IPs (with spin periods $\sim$ 1000 s), because 1000 s $\times$ $R$ $\sim$ 4 h is 
comparable to their orbital periods ($P_{orb}$)\footnote{More generally, as 
pointed out by Barrett, O'Donoghue \& Warner (1988), the most frequent kind of IP 
has $P_{orb} \approx 14 P_{spin}$, which we see results in $P_{QPO} \approx P_{orb}$. 
There is no obvious physical reason for this coincidence.}, and a QPO signal would then 
need very long observational runs to distinguish it from orbital modulation. 
In GK Per, which has $P_{orb}$ = 2 d, no such confusion arises.

\subsection{RX And}

     Szkody (1976) found 35.7 s DNOs about 1 mag down from the peak of 
an outburst. In the light curve there is some weak evidence for QPOs with a 
period $\sim$ 17 min (see Figure 2 of Szkody (1976)), which gives $R \sim 28$.

\subsection{SW UMa}

    Robinson et al (1987) found DNOs with periods near 22.3 s on several 
nights during the middle and late phases of the superoutburst of SW UMa in 
March 1986. In addition they found QPOs with periods near 300 s. In the 
March 1992 superoutburst, a few days after maximum, Kato, Hirata \& 
Mineshige (1992) found large amplitude QPOs with a period of 366 s, but 
their photometric repetition time was too long (28 s) to be able to detect 
DNOs. We have $R = 13.5$ from the Robinson et al.~observations.

\subsection{U Gem}

    Soft X-Ray modulations have been observed during outbursts of U Gem. 
Those at a typical DNO period, namely 21.1 s, and amplitude of 12\%, were 
found by Cordova et al.~(1980, 1984) in one out of three outbursts observed, 
and a QPO, also of 12\% amplitude, at 585 s in a different outburst (Cordova 
\& Mason 1984. These give $R = 27.8$, but may refer to very different stages 
in the evolution of the outbursts.  DNOs with periods $\sim$ 25 s have also been 
observed with EXOSAT (Mason et al.~1988) and with EUVE during outburst (Long et al.~1996).

\subsection{V533 Her and BT Mon}

    Rodriguez-Gil \& Martinez-Pais (2002) have found 23.33 min quasi-
periodic flaring in the emission lines of the old nova V533 Her. They 
propose that V533 Her could therefore be a candidate intermediate polar, as 
has been suggested also for another old nova, BT Mon, which shows $\sim$ 30 
min quasi-periodic flaring (Smith, Dhillon \& Marsh 1998). However, for 
several years V533 Her had a 63.63 s coherent luminosity variation 
(Patterson 1979) that put it clearly in the DQ Herculis class of rapid rotation 
intermediate polars\footnote{The disappearance was at one time thought to be uncharacteristic of a 
rotational modulation and could 
imply non-radial oscillation of the primary (Robinson \& Nather 1983), but time moves on and (a) no 
mono-periodic white dwarf pulsator has ever been found, (b) no pulsating white dwarf has ceased pulsating,  
(c) the only CV with an oscillating primary is GW Lib (Warner \& van Zyl 1998) in which the mass accretion 
rate is so low that the primary is cool enough to sit in the DA instability strip, and (d) the high mass 
accretion rate in an old nova such as V533 Her produces an accretion disc luminosity that prevents 
detection of flux direct from the primary (but, as in DQ Her and some nova-like variables, does allow the 
reprocessed radiation from the magnetically-channelled accretion to be detected).}. 
It is more probable, therefore, that the QPO flares in 
V533 Her and BT Mon are connected with the same physical phenomenon 
that causes the brightness QPOs. For V533 Her, $R = 23.33/1.060$ = 22.0.

\section{New observations}

We have used the University of Cape Town CCD Photometer (O'Donoghue 1995) on the 30-in, 
40-in and 74-in telescopes of the South African Astronomical Observatory at Sutherland to observe
the CVs listed in this section. A log of these observations is given in Table 1. Many runs
that do not contain oscillations have been omitted from this list.

\begin{table*}
 \centering
  \caption{Observing log.}
  \begin{tabular}{@{}llrlrrrcc@{}}
 Object       & Type         & Run No.  & Date of obs.          & HJD of first obs. & Length    & $t_{in}$ & Tel. &  V \\
              &              &          & (start of night)      &  (+2450000.0)     & (h)       &     (s)   &      & (mag) \\[10pt]
{\bf CR Boo}  & AM CVn       & S6852    & 2003 Mar 08 & 2707.42816  &   5.48      &        6   &  30-in & 14.6 \\
              &              & S6862    & 2003 Mar 10 & 2709.42381  &   5.53      &  6, 7, 8   &  30-in & 14.9 \\[5pt]
{\bf OY Car}  & DN           & S6488    & 2002 Feb 16 & 2323.50036  &   2.88      &        7   &  74-in & 16.1$^{a,b}$ \\
              &              & S6722    & 2003 Feb 01 & 2672.36465  &   2.80      &     4, 5   &  40-in & 12.7$^a$\\
              &              & S6724    & 2003 Feb 01 & 2672.54450  &   1.94      &        5   &  40-in & 12.8$^a$\\[5pt]
{\bf V803 Cen}& AM CVn       & S6731    & 2003 Feb 02 & 2673.55791  &   1.57      &        2   &  40-in & 13.6 \\
              &              & S6735    & 2003 Feb 03 & 2674.46651  &   2.12      &        4   &  40-in & 13.8 \\
              &              & S6740    & 2003 Feb 04 & 2675.56672  &   1.79      &        6   &  40-in & 13.8 \\[5pt]
{\bf Z Cha}   & DN           & S6061    & 2000 Feb 06 & 1581.35872  &   3.93      &        5   &  40-in & 13.6$^a$\\[5pt]
{\bf AQ Eri}  & DN           & S6159    & 2000 Dec 27 & 1906.31477  &   3.01      &        8   &  74-in & 17.5 \\
              &              & S6510    & 2002 Aug 28 & 2515.64338  &   0.66      &     2, 4   &  74-in & 13.4$^b$\\
              &              & S6516    & 2002 Aug 30 & 2517.56521  &   2.14      &     3, 5   &  74-in & 13.5 \\
              &              & S6520    & 2002 Aug 31 & 2518.59417  &   1.89      &     3, 5   &  74-in & 13.5 \\[5pt]
{\bf VW Hyi}  & DN           & S6133    & 2000 Aug 28 & 1785.54592  &   2.84      &     1, 2   &  74-in & 12.0$^c$ \\[5pt]
{\bf WX Hyi}  & DN           & S6248    & 2001 Sep 22 & 2175.51773  &   0.55      &        6   &  40-in & 14.5 \\
              &              & S6463    & 2002 Jul 10 & 2466.48576  &   3.30      &        6   &  30-in & 14.8 \\[5pt]
{\bf TU Men}  & DN           & S6695    & 2002 Dec 26 & 2635.27795  &   4.40      & 1.5, 5, 6  &  74-in & 14.9  \\[5pt]
{\bf CN Ori}  & DN           & S6702    & 2002 Dec 29 & 2638.29558  &   1.71      &        5   &  74-in & 12.3$^d$ \\[5pt]
{\bf HX Peg}  & DN           & S6475    & 2002 Jul 13 & 2469.55275  &   1.06      &        8   &  30-in & 15.8 \\
              &              & S6584    & 2002 Oct 14 & 2562.24520  &   1.55      &        5   &  40-in & 13.1 \\
              &              & S6646    & 2002 Nov 25 & 2604.26516  &   1.95      &     6, 8   &  40-in & 13.1 \\
              &              & S6650    & 2002 Nov 26 & 2605.26913  &   3.04      &        6   &  40-in & 13.1 \\
              &              & S6656    & 2002 Nov 29 & 2608.26764  &   1.62      &     5, 6   &  40-in & 13.2:\\[5pt]
{\bf VZ Pyx}  & DN           & S6066    & 2000 Mar 07 & 1611.29893  &   1.65      &     3, 4   &  74-in & 12.2:\\
              &              & S6569    & 2002 Jun 03 & 2556.59832  &   0.80      &        6   &  40-in & 12.2:\\[5pt]
{\bf V893 Sco}& DN           & S6095    & 2000 Jun 01 & 1697.32730  &   7.12      &        5   &  74-in & 14.1$^a$ \\[5pt]
{\bf EC\,21178-5417}  & NL   & S6544    & 2002 Sep 07 & 2525.28604  &   6.11      &  5, 6, 10  &  40-in & 13.7$^a$ \\
                      &      & S6548    & 2002 Oct 01 & 2549.30417  &   5.27      &     5, 6   &  40-in & 13.6$^a$ \\
                      &      & S6549    & 2002 Oct 03 & 2551.27063  &   2.34      &        6   &  40-in & 13.6$^a$ \\
                      &      & S6551    & 2002 Oct 04 & 2552.27894  &   3.72      &        5   &  40-in & 13.7$^a$ \\
                      &      & S6553    & 2002 Oct 05 & 2553.35755  &   0.80      &        5   &  40-in & 13.8$^a$ \\
                      &      & S6555    & 2002 Oct 06 & 2554.23099  &   1.39      &        6   &  40-in & 13.6$^a$ \\
                      &      & S6557    & 2002 Oct 07 & 2555.22744  &   1.59      &    6, 10   &  40-in & 13.6$^a$ \\
                      &      & S6564    & 2002 Oct 08 & 2556.22808  &   1.95      &        6   &  40-in & 13.6$^a$ \\
                      &      & S6570    & 2002 Oct 09 & 2557.28798  &   2.53      &        5   &  40-in & 13.7$^a$ \\
                      &      & S6574    & 2002 Oct 10 & 2558.33980  &   1.53      &        6   &  40-in & 13.6$^a$ \\
                      &      & S6580    & 2002 Oct 12 & 2560.30188  &   1.47      &        5   &  40-in & 13.8$^a$ \\
                      &      & S6599    & 2002 Oct 31 & 2579.25335  &   0.88      &     4, 5   &  40-in & 13.7$^a$ \\
                      &      & S6634    & 2002 Nov 17 & 2596.25676  &   0.80      &     2, 6   &  40-in & 13.7$^a$ \\
                      &      & S6639    & 2002 Nov 18 & 2597.25077  &   0.62      &       10   &  40-in & 13.8$^a$ \\
                      &      & S6641    & 2002 Nov 24 & 2603.27788  &   1.62      &        6   &  40-in & 13.7$^a$ \\
                      &      & S6660    & 2002 Nov 30 & 2609.26322  &   1.03      &        6   &  40-in & 13.7$^a$ \\
                      &      & S6666    & 2002 Dec 01 & 2610.26547  &   1.06      &        6   &  40-in & 13.6$^a$ \\
                      &      & S6670    & 2002 Dec 17 & 2626.28606  &   0.83      &        6   &  40-in & 13.8$^a$ \\[5pt]

\end{tabular}
{\footnotesize 
\newline 
Notes: DN = Dwarf Nova; NL = Nova-like; $t_{in}$ is the integration time; `:' denotes an uncertain value; $^a$ mean magnitude
out of eclipse; $^b$ B magnitude; $^c$ I magnitude; $^d$ magnitude estimate from VSNET. \hfill}
\label{tab1}
\end{table*}

\subsection{VZ Pyx}

\begin{figure}
\centerline{\hbox{\psfig{figure=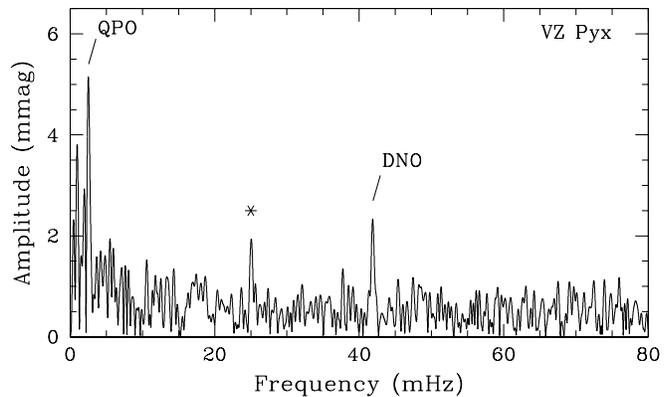,width=8.8cm}}}
  \caption{The Fourier Transform of run S6569 of VZ Pyx. The QPO and DNO periods
are marked. The asterisk marks a spurious signal: the telescope's periodic drive 
error at 40 s.}
 \label{ft6569}
\end{figure}

    VZ Pyx is the optical equivalent of the X-Ray source 1H 0857-242, 
identified by Remillard et al.~(1994) as a CV with dwarf nova outbursts. 
These authors found photometric modulations with periods near 49 mins 
during the decay phases of three outbursts and suggested that this could be 
the signature of an IP. They found no periodic modulation in the X-Ray 
observations (but a possible X-Ray period is quoted by Patterson (1994), the 
evidence for which has not been published) and reported no optical 
modulations at typical DNO periods. Barwig, Wimmer \& Hughes (1993) 
found no periodic modulations at quiescence. VZ Pyx is listed as a probable 
IP in the review by Patterson (1994). On the other hand, Kato \& Nogami 
(1997) point out that its outburst and superoutburst behaviours are like those 
of normal non-magnetic SU UMa stars and not those of outbursting IPs, and 
Szkody \& Silber (1996) find that IUE spectra obtained in quiescence show 
the low excitation emission lines typical of a dwarf nova with a disc and that 
``there is no compelling argument for magnetic accretion from the UV data''.

    We have observed VZ Pyx near maximum of outburst, as reported in 
Table 1, and find both QPOs and DNOs present in run S6569.
The FT of S6569 is shown in Fig.~\ref{ft6569}, where a 
QPO with period 390.5 s and amplitude 5 mmag, and a DNO with mean 
period 23.86 s (the period increased slightly during the run) and mean 
amplitude 2.5 mmag are evident. The ratio $R = 16.4$.

    The QPO is at a much shorter period than the $\sim$ 3000 s modulation seen 
previously near the ends of outbursts. The latter is very long for a QPO in a 
short period ($P_{orb}$ = 1.78 h) system, and is possibly associated with 
inhomogeneities rotating at the outer edge of the disc. In the LIMA model of 
DNOs (Paper II) it is not possible for DNOs of variable period seen directly 
from the primary
to exist in an IP (if the magnetic field is strong enough to 
produce an IP it will prevent free circulation of an equatorial belt) and to 
date none have been observed in any certified 
IP\footnote{Reprocessed DNOs, in which the variable period of the travelling wave is involved, 
may be variable, as in GK Per.}. We conclude that VZ Pyx 
is not an IP. More photometric studies of VZ Pyx late in outburst are 
required to see if the $\sim$ 3000 s modulation is a regular feature and whether 
there is any evidence that it is coherent -- and also to follow the period 
increase in the DNOs.

\subsection{EC\,21178-5417}

    EC\,2117-54 is a V $\sim$ 13.7, UV-rich star discovered in the ongoing 
Edinburgh-Cape Survey (Stobie et al.~1997). We have found it to be an 
eclipsing nova-like CV with $P_{orb}$ = 3.708 h; details of its position, finding 
chart, orbital light curve and spectra will be reported elsewhere -- here we 
describe the results of high speed photometry. Our photometric runs on EC 
2117-54 are listed in Table 1.

    EC\,2117-54 turns out to be a rich source of DNOs and QPOs, and as a 
bright CV with a similarly bright reference star nearby it provides a system 
that is easily studied even in quite poor observing conditions. Its eclipses 
will be particularly useful for investigating phase shifts and physical 
locations of the oscillating sources within the system. In this paper we will 
discuss only the presence and periods of the DNOs and QPOs, and in 
particular in this section we consider only the normal DNOs and QPOs that 
lead to estimates of the ratio $R$. As with other nova-likes (e.g. UX UMa: 
Warner \& Nather 1972; Knigge et al.~1998) the DNOs are not always visible 
in the FTs -- our success rate (at least once in a run) for detecting them in EC 
2117-54 is 83\%.

\begin{table*}
 \centering
  \caption{DNOs, lpDNOs and QPOs in EC\,2117-54.}
  \begin{tabular}{@{}lcrrrrrr@{}}
 Run No.   & Sect. & Length         & \multicolumn{2}{c}{DNOs}  & \multicolumn{2}{c}{lpDNOs}    & \multicolumn{1}{c}{QPOs} \\
           &       & (s)            & \multicolumn{2}{c}{(periods in seconds)} & \multicolumn{2}{c}{(periods in seconds)} & (period in seconds)\\
           &       &                & \multicolumn{2}{c}{[amplitude in mmag]} & \multicolumn{2}{c}{[amplitude in mmag]} & [amplitude in mmag] \\[10pt]
S6544      & I     &  2765          & 23.27 (0.03) & 22.10 (0.04)   & 94.21 (0.26) &     &  480     \\
           &       &                & [2.4]     & [1.5]       & [4.3]     &     &  [3.7] \\
           & II    &  1728          & 23.21 (0.07) &                &  --          &     &   --      \\
           &       &                & [2.2]     &                &              &     &           \\
           & III   &  1728          &   --         &                &  --          &     &   --      \\
           & IV    &  1166          &   --         &                & 96.01 (1.11) &     &    --     \\
           &       &                &              &                & [3.9]     &     &           \\
           & V     &  3240          & 25.46 (0.07):&                & 79.90 (0.43) & 184.0 (1.6) &    \\
           &       &                & [1.9]     &                & [4.2]     & [4.2]    &    \\
           & VI    &  1417          & 23.14 (0.06) &                &   --         &     &  496       \\
           &       &                & [3.3]     &                &              &     &  [7.2]  \\
S6548      & I     &  1797          & 22.43 (0.06) &                &   --         &     &  --        \\
           &       &                & [3.0]     &                &              &     &            \\
           & II    &  2938          &   --         &                &   --         &     &   --       \\
           & III   &  2246          &   --         &                &   --         &     &   --       \\
           & IV    &  1728          &   --         &                &   --         &     &   --       \\
           & V     &  2048          &   --         &                &   --         &     &   --       \\
S6549      &       &  8433          &   --         &                &   --         &     &    --      \\
S6551      & I     &  1210          & 22.54 (0.05) &                & 91.57 (0.10) &     &   --       \\
           &       &                & [2.9]     &                & [3.5]     &     &            \\
           & II    &  1987          & 22.69 (0.05) &                &  --          &     &  500       \\
           &       &                & [2.3]     &                &              &     &  [6.3]  \\
           & IIIa  &  1331          & 10.82 (0.03) & 11.40 (0.03)   & 92.25 (0.57) &     &  500       \\
           &       &                & [4.0]     & [3.9]       & [5.5]     &     &  [6.3]  \\
           & IIIb  &  1365          & 22.80 (0.07) &                & 95.62 (0.80) &     &   --       \\
           &       &                & [2.8]     &                & [4.0]     &     &            \\
           & IIIc  &  1382          & 22.72 (0.04) &                &   --         &     &   --       \\
           &       &                & [3.6]     &                &              &     &            \\
           & IV    &  1296          &   --         &                &   --         &     &  --        \\
           & V     &  1037          &   --         &                &              & 195.1 (2.8) & -- \\
           &       &                &              &                &              & [6.4]    &    \\
S6553      & I     &  1443          &   --         &                &    --        &      &     --     \\
           & II    &  1434          & 23.23 (0.10) &                &    --        &       &    --     \\
           &       &                & [2.0]     &                &              &      &            \\
S6555      &       &  2454          & 12.74$^*$ (0.04) &             &   --        &      &  --        \\
           &       &                & [2.1]     &                &              &      &            \\           
S6557      &  I    &  1953          &   --         &                &    --        &      &  --        \\
           &  II   &  1555          &   --         &                &    --        &      &  --        \\ 
           &  III  &  1987          & 22.68 (0.04):&                &    --        &      &  --        \\
           &       &                & [2.3]     &                &              &      &            \\
S6564      &  I    &   821          &   --         &                &              & 169.5 (2.5) & --  \\ 
           &       &                &              &                &              & [7.6]    &     \\
           &  II   &  2246          &   --         &                &    --        &      & --        \\
           &  III  &  1210          &   --         &                &    --        &      & --        \\
S6570      &  I    &  3197          & 22.67 (0.04) &                &    --        &      & 332         \\
           &       &                & [1.9]     &                &              &      & [7.0]    \\
           &  II   &  1382          &   --         &                &    --        &      &  --        \\
S6574      &  I    &  1987          &   --         &                &    --        &      &  --        \\
           &  II   &  2246          &   --         &                &    --        &      &  --        \\
S6580      &  I    &  1728          & 23.02 (0.05) &                &    --        &      & 373         \\
           &       &                & [3.0]     &                &              &      & [6.5]    \\
           &  II   &  2367          &   --         &                &    --        &      &  --         \\
S6599      &  I    &  1469          & 23.67 (0.11) &                &    --        &      &  --         \\
           &       &                & [2.7]     &                &              &      &             \\
           &  II   &  1728          & 23.41 (0.10) & 11.01 (0.02)   &    --        &      &  --         \\
           &       &                & [2.0]     & [2.5]       &              &      &             \\
S6634      &       &  2868          & 23.28 (0.05) & 13.2$^{**}$      &  --          &      &  --         \\
           &       &                & [2.3]     &                &              &      &             \\[5pt]
\end{tabular}
{\footnotesize 
\newline 
$^*$ This is the observed period, which is probably the beat period with the integration time. The true period
would be 11.31 s.\hfill \\
$^{**}$ This is the observed period, which is probably the beat period with the integration time. The true period
would be 11.0 s.\hfill \\
}
\label{tab2}
\end{table*}
\addtocounter{table}{-1}

\begin{table*}
 \centering
  \caption{continued: DNOs, lpDNOs and QPOs in EC\,2117-54.}
  \begin{tabular}{@{}lcrrrrrr@{}}
 Run No.   & Sect. & Length         & \multicolumn{2}{c}{DNOs}  & \multicolumn{2}{c}{lpDNOs}    & \multicolumn{1}{c}{QPOs} \\
           &       & (s)            & \multicolumn{2}{c}{(periods in seconds)} & \multicolumn{2}{c}{(periods in seconds)} & (period in seconds)\\
           &       &                & \multicolumn{2}{c}{[amplitude in mmag]} & \multicolumn{2}{c}{[amplitude in mmag]} & [amplitude in mmag] \\[10pt]
S6639      &       &  2229          & 23.18 (0.09) &                &  --          & \ \ \ \ \ \ \ \ \ \ \ \     &  --      \\
           &       &                & [3.3]     &                &              &     &           \\
S6641      &  I    &  2920          & 23.18 (0.05) &                & 95.82 (0.69) &     &    --     \\
           &       &                & [2.1]     &                & [2.7]     &     &           \\
           &  II   &  2773          &  --          &                & 101.03 (0.55): &    &  --      \\
           &       &                &              &                & [2.9]       &    &          \\
S6660      &  I    &  778           &  --          &                &  --            &    & --         \\
           &  II   &  1166          & 23.19 (0.11) & 21.42 (0.12)   &  --            &    & 214:         \\
           &       &                & [2.6]     & [2.0]       &                &    & [7.9]     \\
S6666      &  I    & 1909           &  --          &                &   --           &    &    --      \\
           & II    & 1909           & 23.29 (0.05) &                &    --          &    &    --       \\
           &       &                & [2.8]     &                &                &    &           \\
S6670      &       & 1840           &    --        &                &    --          &    &    --      \\[5pt]
\end{tabular}
{\footnotesize 
\newline 
Notes: Uncertainties in the periods are quoted after the periods between brackets;  `:' denotes an uncertain value. \hfill \\
}
\label{tab2b}
\end{table*}

      Table 2 contains the list of detected oscillations (the meaning of lpDNOs 
is given in Section 4). In some cases the variations within a run are 
described in more detail in Section 4.2.2.
The DNO periods cluster around 23 s but the differences are much 
greater than the errors (derived from least squares fits of sinusoids to the 
DNOs), which provides the first indication that these are indeed DNOs and 
not produced by stable rotation of the primary (as, e.g., in DQ Her).   The 
periodicities are not always present at full strength throughout the individual 
runs. 

\begin{figure}
\centerline{\hbox{\psfig{figure=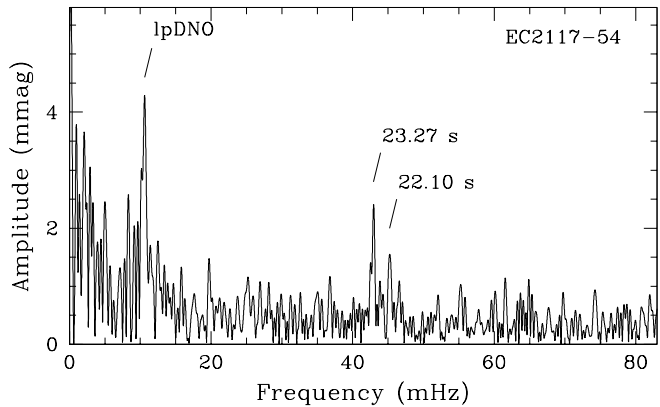,width=8.8cm}}}
  \caption{The Fourier Transform of the first 45 minutes of run S6544 of EC 2117-54. 
The lpDNO and DNO period are marked. The light curve has been prewhitened at the low
frequency modulation visible in Fig.~\ref{lc6544}.}
 \label{ft6544}
\end{figure}

     In run S6544 part (I) there is a double DNO with periods 22.10 s and 
23.27 s -- see Fig.~\ref{ft6544}. These beat together at a period of 440 s; there is a 
broad low amplitude signal in the FT near 470 s (which we would not have 
thought significant without the DNO evidence, but it should be remembered 
that the unstable nature of QPOs spreads and lowers their amplitude in an 
FT). By analogy with VW Hyi and other CVs with double DNOs, the shorter 
of the two periods is the true DNO, the longer period is a sideband formed 
by reprocessing of the revolving DNO beam off the QPO travelling wave 
(Paper II). We therefore have $R = 440/22.1 = 19.9$.

     In part (III) of run S6551 there is a double DNO at 10.82 s and 11.40 s. 
These are the harmonics of fundamentals of which only one is seen in the 
FTs of another portion of the run.  The periods 2 x 10.82 s and 22.69 s beat to 
give a pseudo QPO period of 468 s, but in this case a genuine QPO is seen in 
the FT of that part of the run, centred on 500 s and with amplitude 6.3 
mmag. This run gives $R$ = 21.6.

     In run S6599 part (II) there are two DNOs in the FT -- at 23.41 s and 
11.01 s. The latter is a first harmonic, though again there is no signal at the 
fundamental. The fundamental beats with 23.41 s at a period of 368 s. There 
is no signal at this, putative, QPO period in the FT, but we note that 
368/22.04 gives $R$ = 16.7.

   Run S6660 has a double DNO that has a beat at 281 s; the observed QPO 
is near 214 s. The observed $R$ = 13.1. Runs S6570 and S6580 give observed 
values of $R$ = 14.6 and $R$ = 16.2 respectively.

    Runs S6555 and S6634 contained apparent harmonics that are probably beats 
with the integration lengths. The true periods are given in footnotes to Table 
2.

\subsection{CN Ori}

CN Ori was observed on 2002 December 29 during outburst. A double DNO was visible
in the first $\sim$ 1 hr of the light curve with peaks at 12.10 s and 11.23 s. 
The beat period between these two periodicities is 156.6 s; there is no sign in the
FT of a peak at this period. This run gives $R$ = 13.9.

\subsection{WX Hyi}

\begin{figure}
\centerline{\hbox{\psfig{figure=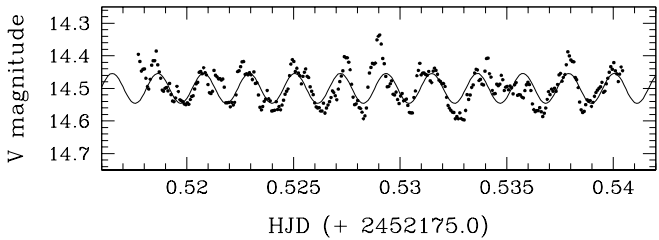,width=8.8cm}}}
  \caption{The light curve of WX Hyi (run S6248), showing the 185-s QPO clearly. 
Superimposed is the result from the non-linear sinusoidal least-squares fit.}
 \label{lc6248}
\end{figure}

 We observed WX Hyi following an outburst in September 2001, when it had
almost returned to quiescence. In this run (S6248) strong QPOs at 185 s are visible
(see Fig.~\ref{lc6248}). These QPOs remained coherent for $\sim$ 10 cycles.
At the same time, DNOs are apparent in the FT at 19.4 s (amplitude 3.8 mmag). This gives
$R$ = 9.5. QPOs at 191 s were also observed during another run of WX Hyi
in July 2002 (run S6463), but will be reported in detail elsewhere.

\subsection{OY Car} 

Optical DNOs with periods from 19.4 s to 28.0 s have been seen in OY Car at the
end of superoutburst (Schoembs 1986), and 18 s DNOs have been observed with the
HST, also at the end of superoutburst (Marsh \& Horne 1998). No simultaneous
DNOs and QPOs have hitherto been detected.

We have extensive coverage of the February 2003 superoutburst of OY Car, but list
in Table 1 only the runs that are relevant to the current discussion, which are near maximum
light and obtained on the same night with an interruption between them.

\begin{figure}
\centerline{\hbox{\psfig{figure=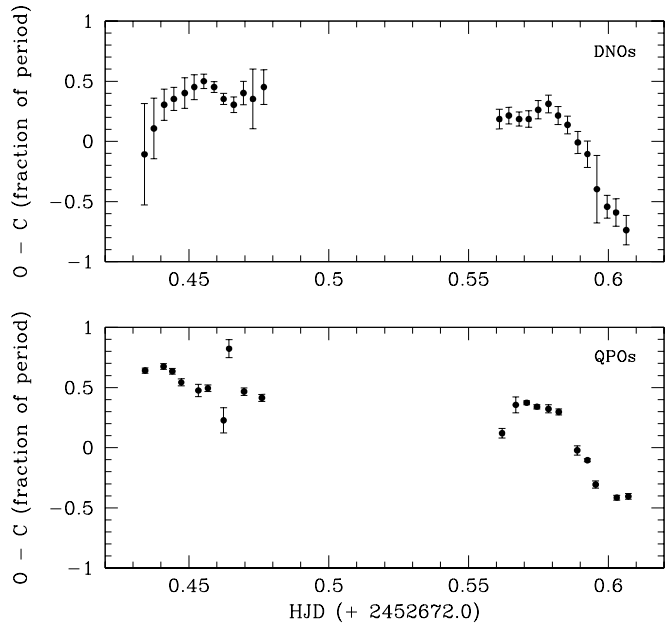,width=8.8cm}}}
  \caption{The O--C diagram of DNOs (upper panel) and QPOs (lower panel) in OY Car.
Data from run S6722 are shown on the left, data from run S6724 are displayed on the right.
The DNOs are displayed relative to a period of 17.62~s (run S6722) and 17.79~s (run S6724); the QPOs
are displayed relative to a period of 281~s (run S6722) and 338~s (run S6724).}
 \label{oc6722}
\end{figure}

Run S6722 shows a DNO at 17.62 s and QPOs at 281 s; later that night run S6724 shows
a DNO at 17.79 s and a QPO initially at 338 s which later changed to 297 s. The phase
diagrams shown in Fig.~\ref{oc6722} show a correlation between the behaviour
of the DNOs and QPOs. Values of $R$ = 15.9 and 19.0 are given by these data.

\subsection{Z Cha}

\begin{figure}
\centerline{\hbox{\psfig{figure=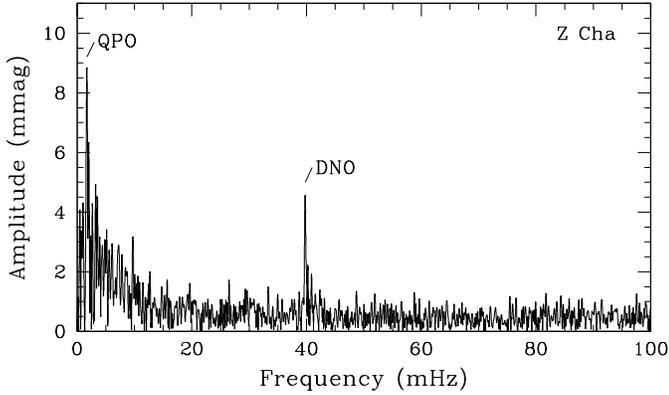,width=8.8cm}}}
  \caption{The Fourier Transform of the first $\sim$ 1.5 hr of run S6061 of Z Cha.
The QPO and DNO periodicities are marked.}
 \label{ftzcha}
\end{figure}

Z Cha was observed during the February 2000 outburst (run S6061, see Table 1)
and a distinct DNO at 25.15 s was seen in the FT. The FT of the prewhitened light curve
(of the first 1.5 hr of run S6061) is shown in Fig.~\ref{ftzcha}. In this section
a distinct QPO signal at 585 s is present. This gives
$R \approx$ 23.3.

\subsection{AQ Eri}

AQ Eri is an SU UMa type dwarf nova with an orbital period of 1.46 h and
a magnitude range V = 12.5 -- 17.5. It was reported at V = 12.9 on 27 August 2002; we
observed it on its decline from maximum. The observations are listed in Table 1. 
DNOs and QPOs were detected for the first time in this star. In addition we have
one earlier light curve obtained in quiescence.

\begin{figure}
\centerline{\hbox{\psfig{figure=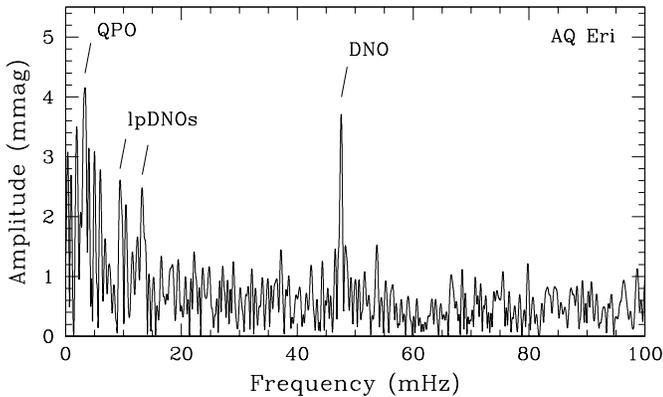,width=8.8cm}}}
  \caption{The Fourier Transform of run S6510 of AQ Eri. The QPO, DNO and lpDNO frequencies
are marked.}
 \label{ft6510}
\end{figure}

The FT of the first light curve (run S6510), with the lowest frequencies 
(partly caused by a superhump) filtered out, is shown in Fig.~\ref{ft6510}. There is
a very prominent single DNO at 21.0 s and a broad QPO feature centred on $\sim$ 296 s.
These give $R \approx$ 15.0. There are other spikes in the FT, which will be discussed
in Sect. 4.7.

\begin{figure}
\centerline{\hbox{\psfig{figure=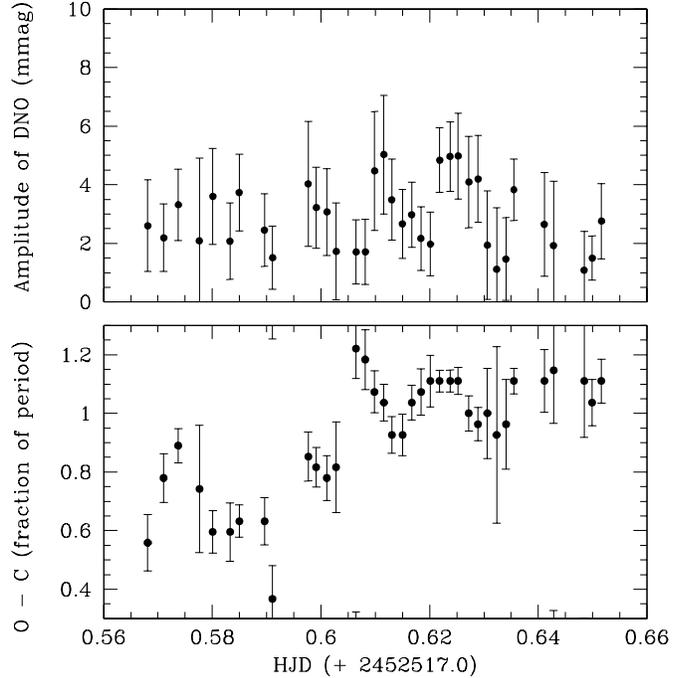,width=8.8cm}}}
  \caption{The O--C diagram (lower panel) of the 23.5 s DNO in AQ Eri (run S6516). The upper panel
shows the amplitude variations of the DNO.}
 \label{oc6516}
\end{figure}

Two nights later (run S6516) the filtered FT again shows a QPO centred on $\sim$ 270 s and
the DNO has lengthened to 23.5 s and now shows considerable structure. The latter is
caused by lower coherence, as illustrated in the O--C diagram (Fig.~\ref{oc6516}). For this
night $R$ = 12.8.

One night later there is no evidence for a QPO or a DNO. The latter probably had become
too incoherent to be detected in an FT.

\subsection{V893 Sco}

\begin{figure}
\centerline{\hbox{\psfig{figure=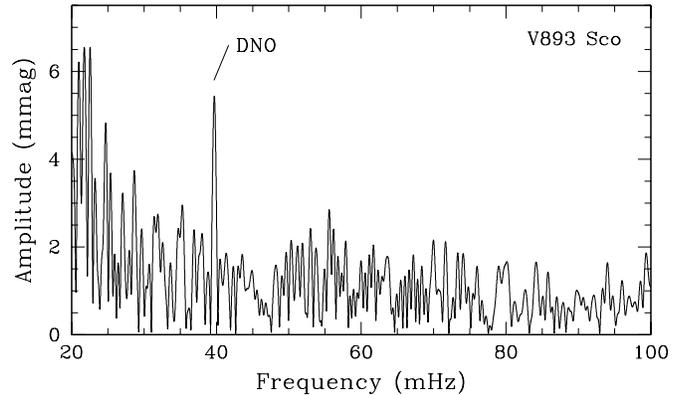,width=8.8cm}}}
  \caption{The Fourier Transform of a section of run S6095 of V893 Sco. The DNO frequency
is marked.}
 \label{ft6095}
\end{figure}

V893 Sco is an eclipsing dwarf nova with an orbital period of 1.82 h. We have observed it 
extensively and find that it frequently has large amplitude QPOs with periods $\sim$ 350 s,
as first detected (at 343 s) in one light curve by Bruch, Steiner \& Gneiding (2000). Only part of one of our
runs (S6095) has a DNO in it. This is listed in Table 1 and was obtained during an outburst.
The FT is shown in Fig.~\ref{ft6095}; the large amount of `red noise' is largely due
to QPO and flickering activity. We removed the eclipses in the light curve before
this analysis. The QPO period is 375 s and the DNO period is 25.2 s, which gives $R$ = 14.9.

\subsection{TU Men}

\begin{figure}
\centerline{\hbox{\psfig{figure=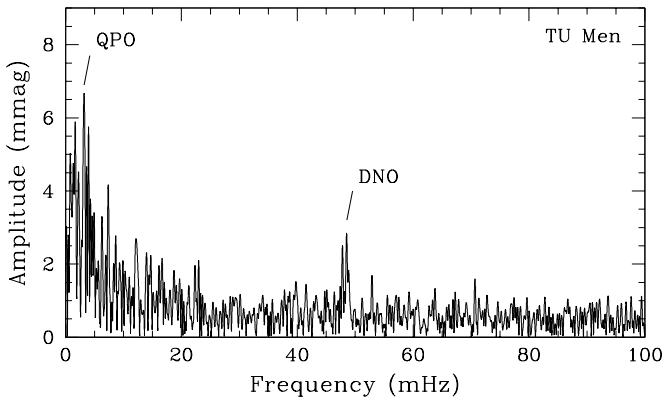,width=8.8cm}}}
  \caption{The Fourier Transform of the first hour of run S6695 of TU Men. Both the QPO and DNO frequencies
are marked.}
 \label{ft6695}
\end{figure}

TU Men is an SU UMa type dwarf nova with an orbital period above the period gap of 2.813 h.
We observed TU Men on decline from outburst (approximately mid way down) on 
2002 December 26 (run S6695, see Table 1), and detected simultaneous
DNOs and QPOs in the first hour of the light curve. The FT of the first hour of run S6695 is shown in Fig.~\ref{ft6695}.
The DNO period is 20.6 s and the QPO period is 313 s, which gives $R$ = 15.2.

\section{Other DNOs/QPOs: A new type of DNO}

    In a few CVs more than one set of DNOs or QPOs has been observed. 
This raises the possibility that if only one modulation is seen it may be 
difficult to know to which group it belongs (see, e.g., Sect. 4.10). 
We therefore examine what is 
known about oscillations other than those that have the relationship $P_{QPO} \sim 
15 P_{DNO}$. We start by discussing observations of VW Hyi.

\subsection{VW Hyi}

     During seven of nine nights (22 -- 30 December) of the superoutburst of 
VW Hyi in December 1975, Haefner, Schoembs \& Vogt (1977, 1979; see 
also Schoembs 1977) saw a persistent sinusoidal modulation with 
amplitudes from 0.5 to 11 mmag, and an apparently erratically changing 
period in the range 86.0 -- 92.8 s (averaging $\sim$ 88.1 s) but unchanging in 
mean period as VW Hyi decreased in brightness from 10.6 to 13.7 mag. This 
contrasts with the rapidly increasing periods of the normal DNOs and QPOs 
over this magnitude range (Paper I). Schoembs \& Vogt (1980) detected a 74 
s QPO near the end of the October 1978 VW Hyi superoutburst. 

\subsubsection{New observations and possible interpretation}

    We have observed similar `longer period' dwarf nova oscillations 
(lpDNOs) in VW Hyi, but only on rare occasions. Of the 51 runs on VW Hyi 
listed in Table 1 of Paper I, lpDNOs are definitely present in only five 
(S0018: 75.6 s; S0129: 93.1 s; the first half of S1594: 71.2 s; S2243: 75.9 s; 
and S6184: 83.0 s), and a run accidentally omitted from Paper I (S6133: see 
Table 1, made near the end of a normal outburst) which contains two 
oscillations, near 85 s and 94 s. These lpDNO detections are distributed from 
maximum light almost to quiescence, and include normal and 
superoutbursts. The oscillation in S2243 is relatively strong (amplitude 
1.3 mmag) and is from early in the same December 1975 superoutburst 
that Haefner et al.~saw the  $\sim$ 88 s modulations. The amplitudes of the 
oscillations in S6184 are 1.3 mmag, made when VW Hyi was a factor of ten 
less luminous than in S2243. There is a tendency for the periods in our runs 
to increase through the outburst (from 75.9 s at maximum to 93.1 s at the 
end of outburst), but this may be small number statistics (the Haefner et al 
observations do not show this) and in any case the oscillations come from 
six different outbursts.

    We get a clue to what the $\sim$ 70 -- 93 s periodicities in VW Hyi may be by 
noting that Sion et al.~(2001) have measured $v \sin{i}$ = 500 km s$^{-1}$ and $\log{g}$ = 
8.0 for the primary of VW Hyi, and that Sion et al.~(1996) found $v \sin{i}$ = 300 
km s$^{-1}$ from a different set of spectra. With an inclination $i = 65^\circ$ and the 
standard white dwarf mass/radius relationship these give $M(1)$ = 0.6 M$_{\odot}$
and a rotation period $P_{rot} \sim 120 - 165$ s for the primary. The observed periods 
may therefore be associated with the bulk rotation of the white dwarf --
which is certainly a short period clock available in the system. This suggests 
that some mass accretion onto two magnetic poles on the primary is at times 
taking place (as well as onto the belt). Unlike an IP, in a low field CV the 
angular momentum accreting at the equator will be distributed in latitude as 
well as radially; the outer layers of the primary will therefore rotate 
differentially and it follows that the accretion regions may not always have 
exactly the same angular velocity. 

\subsubsection{Phase and amplitude diagrams}

\begin{figure}
\centerline{\hbox{\psfig{figure=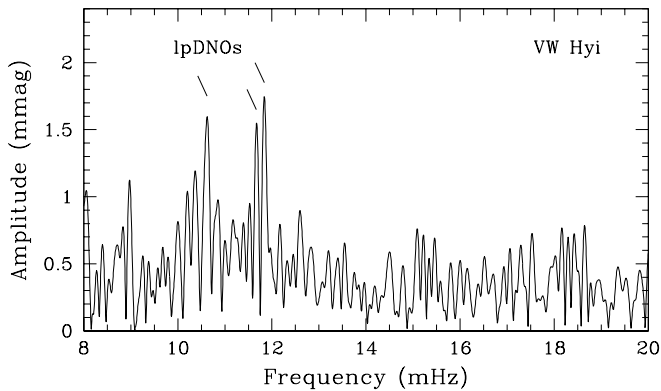,width=8.8cm}}}
  \caption{The Fourier Transform -- truncated at the lowest frequencies --
of VW Hyi (run S6133). The lpDNO frequencies are marked.}
 \label{ft6133}
\end{figure}

     To investigate these lpDNOs more thoroughly we have computed FTs for 
subsets of the data and have also produced phase/amplitude diagrams. 
In such diagrams the formal errors, derived from least squares fits, are sometimes
very large. This is particularly the case where there are large amplitude
QPOs or flickering which has not been filtered out.
We start by describing the behaviour of VW Hyi in run S6133. The FT in the 
vicinity of 100 s for the entire run, shown in Fig.~\ref{ft6133}, has three strong 
periodicities. Using prewhitening and multiple sinusoidal non-linear least 
squares fitting we find these are best represented by oscillations at 84.44 s 
and 85.70 s, with an amplitude of 1.7 mmag and 1.5 mmag, respectively, and 94.13 s with amplitude 1.6 
mmag. The doublet at 85 s is strong only in the first half of the run, which is 
not long enough to resolve it fully; in the second half of the run it appears as 
a single peak at 85.3 s. The separation of the doublet is therefore not well 
determined; although formally it corresponds to a beat period of 5740 s, we 
found periods up to 6200 s while analysing the data in various ways. The 
orbital period of VW Hyi is 6173 s, so we suggest, in analogy with IPs, that 
the 85 s doublet consists of a rotating beam with a period of 84.4 s and an 
orbital sideband produced by reprocessing by the secondary or by the 
vertical thickening of the disc where the stream impacts.

     If the 94 s modulation is also a reprocessed signal then, by analogy with 
the model for double DNOs, it implies a prograde travelling wave with a 
period of 820 s. There is no peak in the FT at this period, but shadowing of 
the disc and shadowing in our direction can be very different for a system 
that is not of high inclination.

\begin{figure}
\centerline{\hbox{\psfig{figure=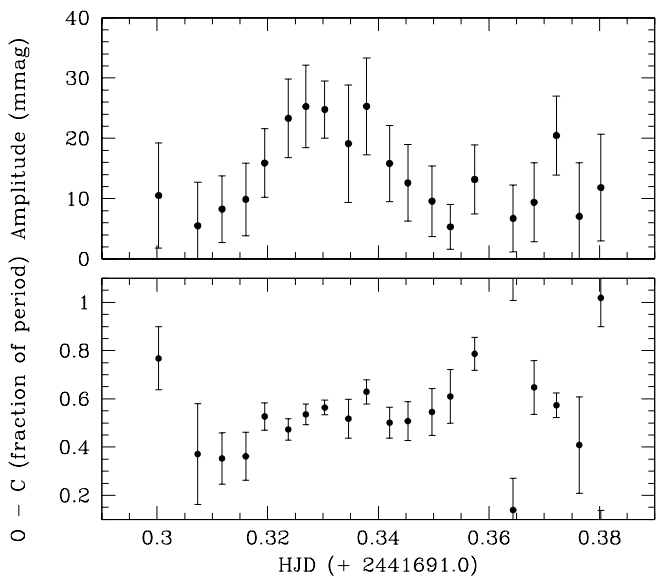,width=8.8cm}}}
  \caption{Amplitude/phase diagrams for the lpDNO in run S0129 of VW Hyi. For the lpDNO
a mean period of 93.1 s is taken, and each dot represents $\sim$ 7 lpDNO cycles; there is a 50\% overlap.}
 \label{omc0129}
\end{figure}

   Of the other runs with lpDNOs in them all but one have narrow spikes in 
their FTs. The 93 s peak in the FT of run S0129 is too broad to be a single 
period, and its phase/amplitude diagram shows a sinusoidal modulation 
suggesting the presence of two unresolved periodicities beating together --
see Fig.~\ref{omc0129}. The sine wave fitted to the amplitude modulation has a period $\sim$
5100 s, but is uncertain because only one cycle is observed. This is of a 
similar time scale to the doublet separation in run S6133.

    The occasional presence of unresolved doublets in the FT could account 
for some of the spread in the lpDNO periods seen in VW Hyi by Haefner et 
al.

\subsection{EC 2117-54}

\subsubsection{DNOs}

\begin{figure}
\centerline{\hbox{\psfig{figure=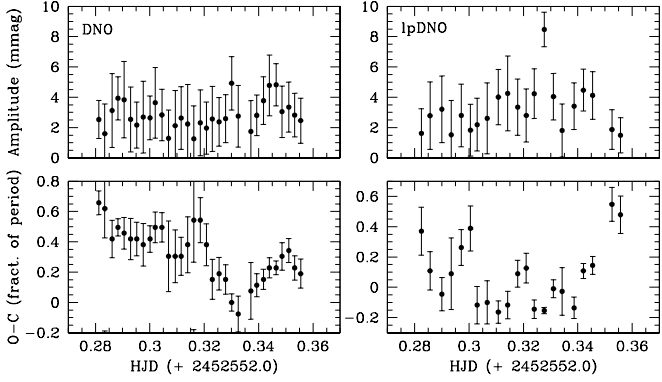,width=8.8cm}}}
  \caption{Amplitude/phase diagrams for the DNO (left panels) and lpDNO (right panels) in run S6551 (Sections I, II and III, see Table 2)
of EC\,2117-54. For the DNO a mean period of 22.69 s was taken, and each dot represents $\sim$ 18 DNO cycles. The lpDNO is plotted
against a mean period of 95.5 s, and each dot represents $\sim$ 6 lpDNO cycles. In both cases there is a 50\% overlap.}
 \label{omc6551}
\end{figure}

     In Section 3.2 we gave the first evidence that the $\sim$ 23 s oscillations are 
DNOs. To study this further we have computed amplitude/phase diagrams 
for all of the runs, some of which are shown in Fig.~\ref{omc6551}. The variations 
seen (for example the change of slope in the O--C diagram from negative to 
positive in run S6551 at HJD 2452552.335 (bottom left panel of Fig.~\ref{omc6551}), which is 
caused by a change in period from 22.66 s to 22.74 s) are characteristic of DNOs 
(see Paper I). There is no systematic variation of DNO phase with orbital phase of the kind seen in DQ 
Her (O'Donoghue 1985). Phase changes in the DNO (as illustrated in Fig.~\ref{omc6551}) occur at different times
as phase changes in the lpDNO, and the two oscillations appear to be independent.

    The interpretation of the 11.02 s period in run S6599 given in Section 3.2 
above enables us to explain the existence of the 12.78 s period in run S6555. 
The integration time for this run was 6s, so an 11 s periodicity would lie 
above the Nyquist frequency. If 12.78 s is the beat frequency with the 
integration length (see RU Peg, Section 2.1) then we are detecting a first 
harmonic modulation at 11.31 s, implying a fundamental at 22.62 s, which is 
within the range of $P_{DNO}$ seen in other runs. In the FT there is no sign of the 
fundamental. 

     The same explanation holds in run S6634, where the observed 13.2 s 
modulation of very small amplitude is really an 11.0 s periodicity, implying 
a fundamental at 22.0 s and a pseudo QPO period $\sim$ 400 s. There is nothing 
significant in the FT at this latter period.

\subsubsection{lpDNOs and QPOs}

     It can be seen in Table 2 that, in addition to the regular DNO and QPO 
modulations, a period around 94 s frequently appears in the light curve. For 
example, in addition to its double DNO, run S6544 has a sinusoidal 
modulation at a period near 94 seen in Fig.~\ref{ft6544} (Note that the mean 
amplitude, derived from fitting to a sinusoid, especially one that is not quite 
constant in period (see below), is much smaller than the range of the largest 
cycles, which are the ones most readily seen in the light curve.) 

\begin{figure}
\centerline{\hbox{\psfig{figure=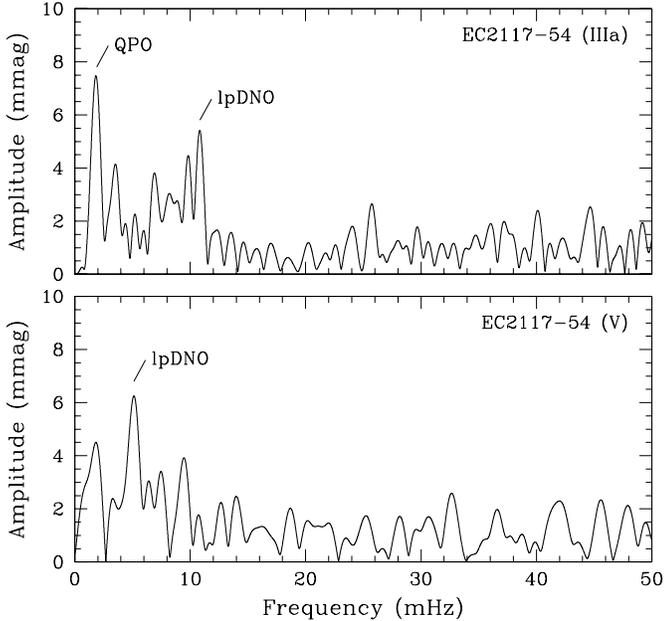,width=8.8cm}}}
  \caption{The Fourier Transform of Section IIIa (upper panel) and V (lower panel)
of run S6551 of EC\,2117-54.}
 \label{ft6551}
\end{figure}

    The lpDNO oscillations are also easily visible in the light curve of run 
S6551, but this run has the added interest that in one section the period 
doubles (Fig.~\ref{ft6551}), which could signify a temporary change from two-pole 
to single-pole accretion.

     In addition, in EC 2117-54 there are regular QPOs with periods $\sim$ 500 s 
that are easily seen in some sections of the light curves.

    The suite of periods in EC 2117-54 is remarkably similar to that seen in 
VW Hyi late in its decline from outburst. In the latter, QPOs with $P_{QPO} \sim 400$ 
s and DNOs with $P_{DNO} \sim 22$ s occur when the system is about 1 mag above 
its quiescent brightness (Paper I). Furthermore, the $\sim$ 94 s periodicities that 
we have seen in EC 2117-54 are similar to those in VW Hyi at $\sim$ 88 s (see 
Section 4.1.1).

\subsubsection{Some detailed examples}

     The richness of the observational material for EC 2117-54 enables us to 
make a detailed analysis of the behaviour of the lpDNOs and compare them 
with the DNOs. As these are a newly recognised class of oscillation we 
present more detail than might otherwise be justified. We will start by 
featuring two light curves where the lpDNOs are particularly prominent. 

\begin{figure}
\centerline{\hbox{\psfig{figure=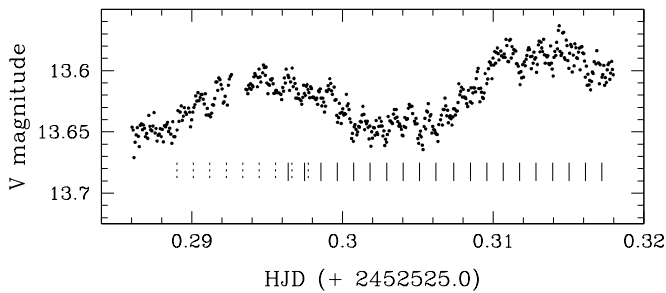,width=8.8cm}}}
  \caption{The light curve EC 2117-54 (the first 45 minutes of run S6544).
The lpDNO modulation at 94.21 s is clearly visible in the light curve.
The lpDNO minima are marked by vertical bars. There is a phase shift around 
HJD $\sim$ 2452525.297.}
 \label{lc6544}
\end{figure}

    Fig.~\ref{lc6544} shows the first 45 mins of the light curve obtained in run 
S6544. The FT of this section of S6544 has already appeared in Fig.~\ref{ft6544} and was discussed above. 
A phase shift around HJD $\sim$ 2452525.297 is present, illustrated by the two sets of vertical bars
marking the lpDNO minima in Fig.~\ref{lc6544}; the dotted bars fit the first part of the light curve well, the solid
bars fit the minima after the phase shift. This illustrates the lack of complete stability in lpDNOs.

\begin{figure}
\centerline{\hbox{\psfig{figure=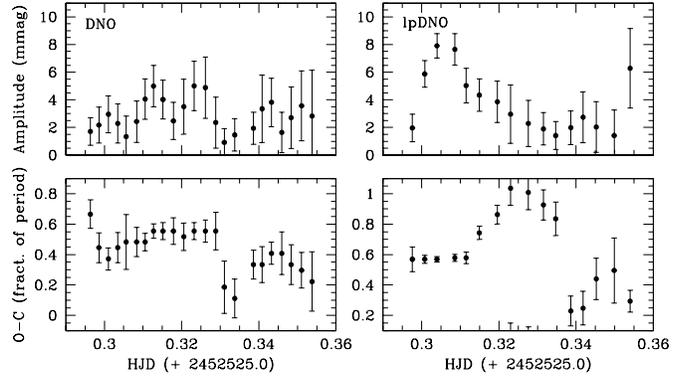,width=8.8cm}}}
  \caption{Amplitude/phase diagrams for the DNO (left panels) and lpDNO (right panels) in run S6544 (Sections I, II and III, see Table 2)
of EC\,2117-54. For the DNO a mean period of 23.27 s was taken, and each dot represents $\sim$ 18 DNO cycles. The lpDNO is plotted
against a mean period of 94.21 s, and each dot represents $\sim$ 6 lpDNO cycles. In both cases there is a 50\% overlap.}
 \label{omc6544}
\end{figure}

Another example of the independent behaviour of the DNOs and lpDNOs is given in
Fig.~\ref{omc6544}, where the amplitude and phase of the DNO and lpDNO in run
S6544 (Sections I -- III) is plotted. 

\subsubsection{Discussion}

     EC 2117-54 has an orbital period that places it in the 3 -- 4 h range where 
most, perhaps all, nova-like variables experience states of low $\dot{M}$ and are 
known as VY Scl stars. DNOs have not been observed in any recognised VY 
Scl stars but they have in two nova-likes that will probably show low states 
if observed sufficiently (HL Aql: $P_{DNO}$ = 19.6 s, V = 13.5 (Haefner \& 
Schoembs 1987); BP Lyn: $P_{DNO}$ = 25.5 s, V = 14.2 (Ringwald 1992)). Thus 
the DNOs and QPOs have yet to be followed into and out of low states -- it 
will be very instructive to observe the evolution of the oscillations during 
these transitions; all three of the above mentioned objects should be bright 
enough in their low states for such observations to be made.

     Some of the nova-likes in this same period range are classified as 
eclipsing SW Sex stars, two of which have been shown to have periodically 
modulated circular polarization -- LS Peg at 29.6 min (Rodriguez-Gil et al.~2001) 
and V795 Her at 19.54 min (Rodriguez-Gil et al 2002) and are IPs. If 
our interpretation of EC 2117-54 is correct, it is a similar system, probably 
with a weaker magnetic field, and with a rotation period (deduced from the 
lpDNOs) of $\sim$ 3.2 min.

\subsection{SS Cyg}                                                  

    Another star for which longer period DNOs have been seen is SS Cyg. 
Patterson (1981; see also Robinson \& Nather 1979) observed what he 
described as QPOs in the range 32 -- 36 s in two different outbursts of SS 
Cyg, occurring on three nights in the September 1978 outburst when there 
were also $\sim$ 10 s normal DNOs present. There have been no other reports of 
modulations near this period. Sion (1999) finds $v \sin{i}$ = 300 km s$^{-1}$ for SS 
Cyg; with a mass of 1.20 M$_{\odot}$ and a (rather uncertain) inclination of 50$^\circ$ 
this gives a rotational period $\sim$ 63 s which again could give the observed 
modulation if there is two-pole accretion.

\subsection{HT Cas}

      Patterson (1981) reported $\sim$ 100 s QPOs in outbursts of HT Cas, which 
has typically $\sim$ 20 s normal DNOs during outbursts. The QPOs are said to be 
often seen also during quiescence, and therefore represent an invariant 
period in the system, unlike the DNOs. We classify them as lpDNOs.

\subsection{AH Her}

 Patterson (1981) also found $\sim$ 100 s QPOs in AH Her, whose normal 
DNOs range from 24 to 39 s. 

      The HT Cas and AH Her periodicities may be further examples of  
lpDNOs, and therefore possibly associated with the rotation of the primary 
(rotational velocities are not available for these two stars).

\subsection{OY Car}

\begin{figure}
\centerline{\hbox{\psfig{figure=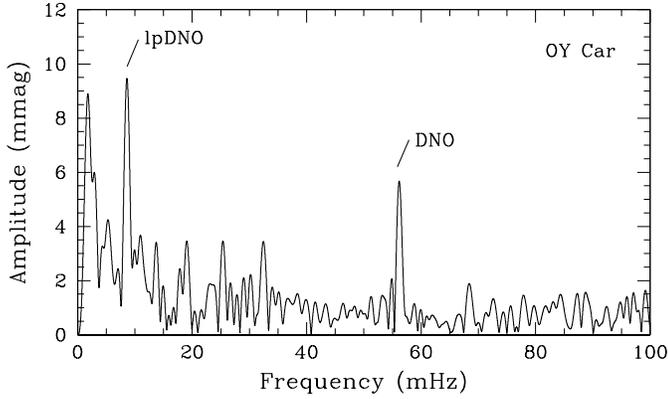,width=8.8cm}}}
  \caption{The Fourier Transform of a section of $\sim$ 20 minutes
of run S6724 of OY Car during superoutburst. The lpDNO and DNO frequencies are marked.}
 \label{ft6724}
\end{figure}

\begin{figure}
\centerline{\hbox{\psfig{figure=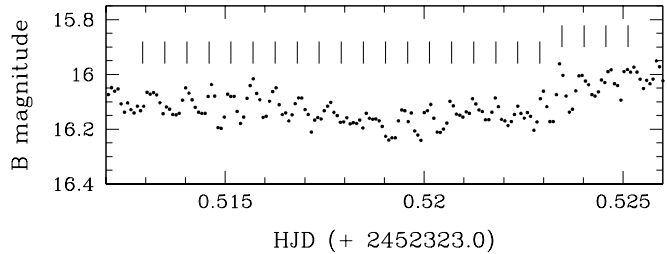,width=8.8cm}}}
  \caption{A section of the light curve of OY Car (run S6488) during quiescence, in which
a clear $\sim$ 48-s periodicity is present (the vertical bars mark the maxima of this coherent
signal).}
 \label{lc6488}
\end{figure}

One section of run S6724 shows a quite coherent modulation with a mean amplitude of 9.5 mmag
and a period of 116 s. The FT is shown in Fig.~\ref{ft6724} and represents ten cycles
of this lpDNO. No other long-lived lpDNOs appear in our light curves.

OY Car has $v \sin i \le$ 200 km s$^{-1}$ (Sion 1999), a mass of 0.68 M$_{\odot}$ and 
$i = 83^{\circ}$ (Wood et al.~1989), which leads to $P_{rot} \ge 260$ s. An lpDNO at
116 s is therefore just compatible with a two-pole accretor.

We have one run (among many others obtained over past years) of OY Car in
quiescence (S6488) which shows a coherent signal. The light curve is given in Fig.~\ref{lc6488} and
shows a period of 47.9 s. This unusual period is perhaps an evolution from the DNOs
seen during outbursts.

\subsection{AQ Eri}

\begin{figure}
\centerline{\hbox{\psfig{figure=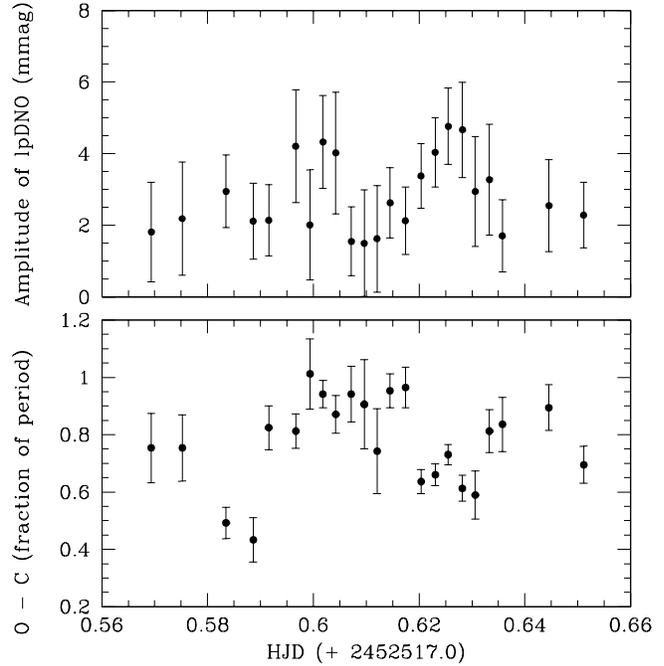,width=8.8cm}}}
  \caption{The O--C diagram (lower panel) of the 73.6 s lpDNO in AQ Eri (run S6516). The upper panel
shows the amplitude variations of the DNO.}
 \label{oc6516b}
\end{figure}

\begin{figure}
\centerline{\hbox{\psfig{figure=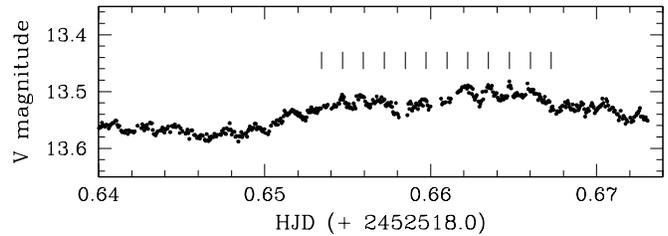,width=8.8cm}}}
  \caption{The light curve AQ Eri on 2002 August 31 (run S6520). 
The lpDNO modulation at $\sim$ 108.6 s is visible in part of the light curve and the most
prominent cycles are marked by the vertical bars.}
 \label{lc6520}
\end{figure}

In addition to the QPOs and DNOs described in Sect. 3.7 we observe lpDNOs in 
AQ Eri. In Fig.~\ref{ft6510} there are two spikes near 100 s which are at
75.7 ($\pm$0.3) s and 106.3 ($\pm$0.5) s with mean amplitudes of 2.5 and 2.6 mmag, respectively 
(derived from non-linear least-squares fits). The frequency difference between these corresponds
to a QPO period of $\sim$ 265 s, which is similar to the observed period (Sect. 3.7).
These appear to be direct and reprocessed lpDNOs as in VW Hyi (Sect. 4.1.2).

In run S6516 a single strong lpDNO at 73.6 ($\pm$0.1) s and mean amplitude of 2.0 mmag
is present. The amplitude/phase plot for this lpDNO is shown in Fig.~\ref{oc6516b}, which
may be compared with that for the DNO that is present simultaneously (Fig.~\ref{oc6516}). 
The lpDNO has greater coherence than the DNO.

The next night, despite the absence of DNOs or QPOs, there is a prominent lpDNO in the
last half of the light curve, being most prominent in the final section as shown in Fig.~\ref{lc6520}. 
The period is 108.6 ($\pm$0.8) s and the mean amplitude is 5.2 mmag, and corresponds to the
reprocessed lpDNO that appears in Fig.~\ref{ft6510} as discussed above.

Our quiescent light curve (S6159) shows large scale flickering but there is a possible lpDNO
of high coherence at 87.0 ($\pm$0.2) s and mean amplitude 7.7 mmag.

\subsection{U Gem}

 The soft X-Ray observations by Cordova, Mason et al., referred to in 
Section 2.13 above, revealed a suite of modulations. In addition to the 21 s 
DNO and 585 s QPO there were observed periodicities at $\sim$ 121 s and $\sim$ 135 
s in quiescence and possibly $\sim$ 152 s in outburst. Robinson \& Nather (1979) 
observed low coherence optical modulations near 146 s and its first 
harmonic during an outburst. These oscillations appear to have the properties 
of lpDNOs, in which case U Gem is the first CV to show all three types of 
oscillation in the X-Ray region.

    Sion et al.~(1998) find $v \sin{i} \sim 100$ km s$^{-1}$ for the primary and a mass of 
1.1 M$_{\odot}$, which with $i$ = 67$^{\circ}$ give $P_{rot} \sim 275$ s, and we again find that two 
pole accretion onto the primary will provide the observed lpDNO periods.

\subsection{VZ Pyx}

\begin{figure}
\centerline{\hbox{\psfig{figure=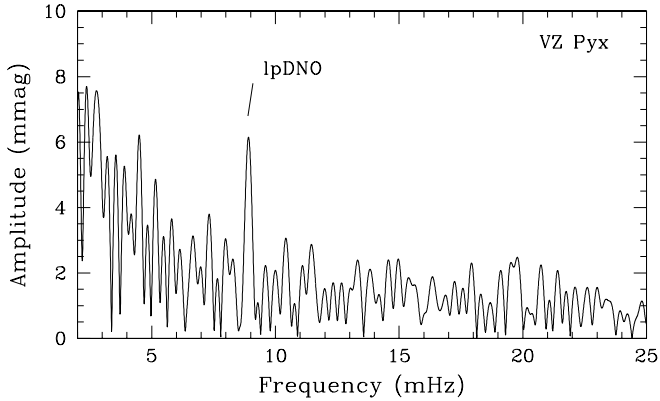,width=8.8cm}}}
  \caption{The Fourier Transform (truncated at the lowest frequencies)
of the last $\sim$ 1 hr of run S6066 of VZ Pyx during outburst. 
The lpDNO frequency is marked.}
 \label{ft6066}
\end{figure}

In the FT of the last half of the run S6066 (see Fig.~\ref{ft6066}), a coherent signal
is present at 112.4 ($\pm$0.6) s, with a mean amplitude 6.2 mmag. No DNOs or QPOs are seen
during this run. Given the suite of DNO and QPO frequencies already identified in Sect.~3.1,
we associate the coherent 112.4-s signal with an lpDNO.

\subsection{HX Peg}

HX Peg is a dwarf nova with an orbital period of 4.82 h and frequent outbursts (recurrence
time $\sim$ 30 d) taking it from a minimum V $\sim$ 16.5 to maximum at V $\sim$ 12.9. We have
observed parts of three outbursts (Table 1).

\begin{figure}
\centerline{\hbox{\psfig{figure=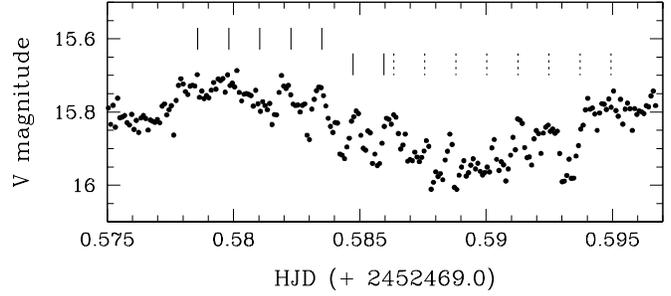,width=8.8cm}}}
  \caption{The light curve of HX Peg (the last $\sim$ 30 minutes of run S6475).
The lpDNO modulation at $\sim$ 112 s is clearly visible; the maxima are marked by vertical bars.
There is a phase shift around HJD $\sim$ 2452469.586.}
 \label{lc6475}
\end{figure}

\begin{figure}
\centerline{\hbox{\psfig{figure=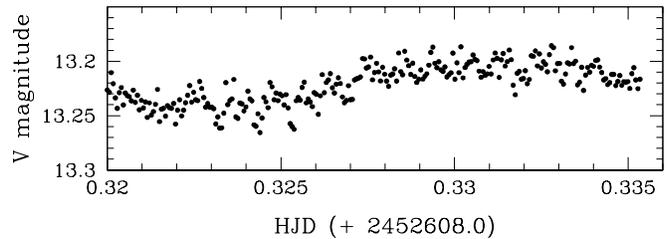,width=8.8cm}}}
  \caption{The light curve of HX Peg (the last $\sim$ 23 minutes of run S6656).}
 \label{lc6656}
\end{figure}

\begin{figure}
\centerline{\hbox{\psfig{figure=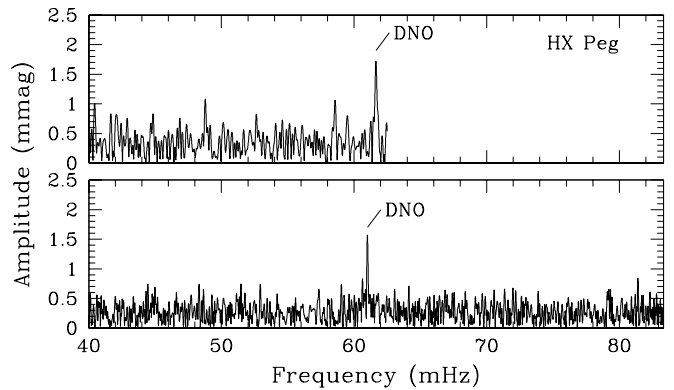,width=8.8cm}}}
  \caption{The Fourier Transform (truncated at the lowest frequencies)
of runs S6646 (upper FT) and S6650 (lower FT) of HX Peg during outburst.
The DNO frequencies are marked.}
 \label{fthxpeg}
\end{figure}

Late in the July 2002 outburst there is a clear signal (run S6475) with a period of 112.3 
($\pm$ 1.1) s and mean amplitude 2.6 mmag, illustrated in Fig.~\ref{lc6475}. On the late rise
of the October 2002 outburst (run S6584) a similar signal is seen in the first half of the
run: 114.8 ($\pm$ 0.7) s with an amplitude of 3.1 mmag. Just after maximum of the November 2002
outburst (run S6656) there is an obvious modulation in the light curve (Fig.~\ref{lc6656}) which the
FT shows is the result of two oscillations, with periods 83.4 ($\pm$ 0.5) s and 111.1 ($\pm$ 1.0) s
with amplitudes of 4.8 and 4.2 mmag, respectively. The beat period between these oscillations
is 335 s and there is a peak in the FT near this period.

The above properties, especially the similarity of the periods near 112 s despite a brightness range
of $\sim$ 2.7 mag (Table 1) suggests an lpDNO, in which the true period is $\sim$ 83 s, and the $\sim$ 112 s
is a reprocessed signal. In none of the runs discussed so far is there any sign of a normal DNO.
However, runs S6646 and S6650, made at the maximum of the November 2002 outburst, have distinct signals
at 16.22 s and 16.39 s, with mean amplitudes of 1.7 mmag and 1.6 mmag, respectively -- seen in the FTs 
in Fig.~\ref{fthxpeg}.
Combined with the QPO observed a few days later in the same outburst (see above) we have $R$ = 20.6.

In addition to the suite of periodicities described above, HX Peg frequently shows very large QPOs
with periods over the range of 1400 -- 1900 s. It was our early observations of these, combined with
the $\sim$ 112 s oscillations that we now recognise to be lpDNOs, that led us to combine
these into the $R$ value $\sim$ 16 announced earlier (Warner \& Woudt 2003). 
We now consider that the $\sim$ 1800 s QPOs arise from a different physical mechanism, perhaps
connected to the rotation period of the outer edge of the accretion disc (see Section 8.7 of
Warner 1995a).

\subsection{AM CVn stars}

\begin{figure}
\centerline{\hbox{\psfig{figure=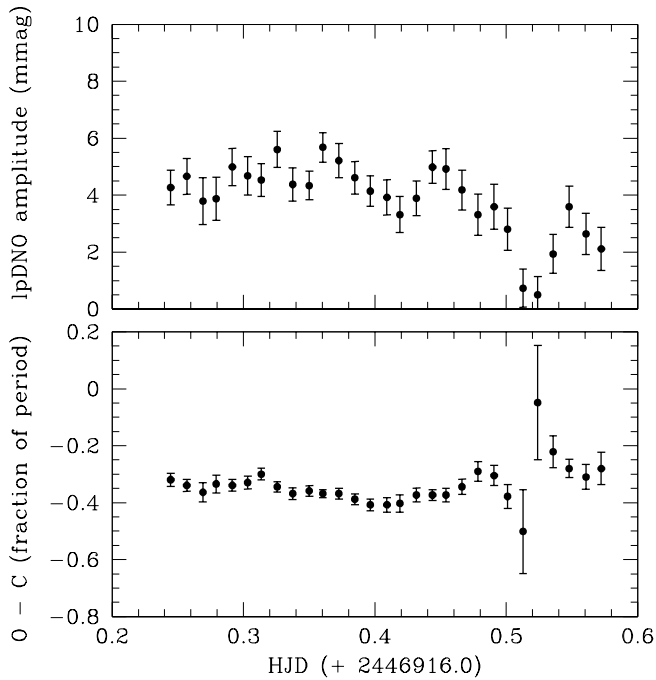,width=8.8cm}}}
  \caption{The O--C diagram (lower panel) and amplitude diagram (upper panel)
of the 176 s photometric modulation seen in V803 Cen (run S4038 of O'Donoghue \& Kilkenny (1989)). }
 \label{omc4038}
\end{figure}

        AM CVn stars are very short orbital period, helium-transferring double 
degenerate CVs. DNOs at 19 -- 22 s and 26.2 s respectively have been 
observed in the high state of CR Boo (Patterson, private communication) and in AM CVn 
(Patterson et al.~1979, Patterson et al.~1992). In most of the AM CVn stars the 
dominant modulation is a non-sinusoidal superhump that has strong 
harmonics to 5th order or higher; this makes the detection of QPOs very 
difficult. Nevertheless, Patterson et al.~(2002) have observed $\sim$ 5 min QPOs 
in HP Lib, which is an AM CVn star in the high state.

\subsubsection{V803 Cen}

\begin{figure}
\centerline{\hbox{\psfig{figure=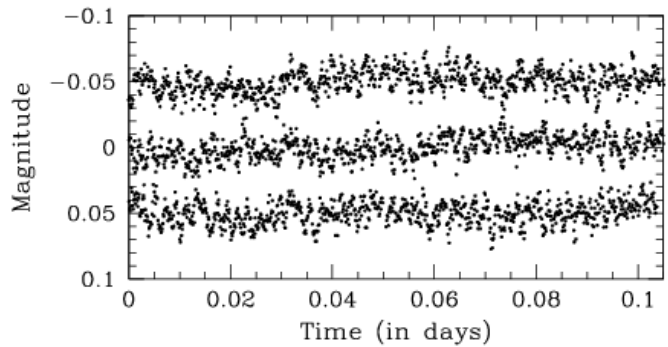,width=8.8cm}}}
  \caption{The light curve of run S4035, prewhitened at the superhump frequency and the first
three harmonics. The data shown represent a continuous light curve of length 0.315 days, split up in three
segments. The first and last segments are displaced vertically, by --0.05 and 0.05 mag, respectively, for display
purposes. The mean magnitude of V803 Cen during this run was B $\sim$ 13.3 mag.}
 \label{lc4035}
\end{figure}

\begin{figure}
\centerline{\hbox{\psfig{figure=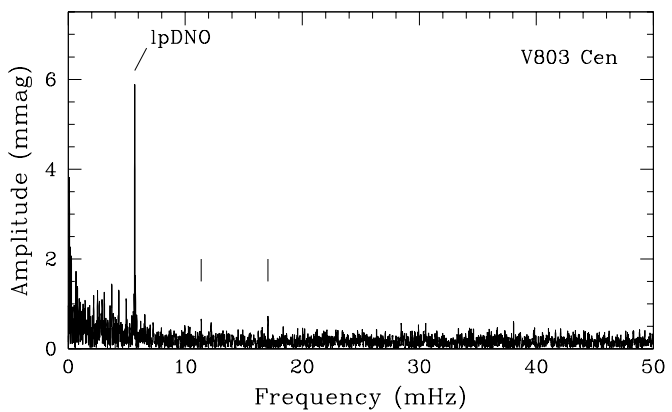,width=8.8cm}}}
  \caption{The FT of run S4035, prewhitened at the superhump frequency and the first
three harmonics. The peak corresponding to the 175.8-s modulation is marked by `lpDNO' and 
the first two harmonics of this modulation are indicated by vertical bars.}
 \label{ft4035}
\end{figure}

      V803 Cen is an AM CVn star with high and low states of mass transfer. 
O'Donoghue, Menzies \& Hill (1987) observed brightness modulations with a mean period 
near 175.8 s during the high state of 2 -- 5 July 1986; the period changed 
only slightly (to 178.8 s) on 7 July when the star had faded by two 
magnitudes. The 176 s modulation was seen again in a high state in 1987 
(O'Donoghue \& Kilkenny 1989), and we have detected it, as described 
below, but it has not been seen in the very low state (O'Donoghue et al.~1990). 
From these observations alone we can already say that the range of 
the oscillations and their insensitivity to luminosity are characteristic of 
lpDNOs.

    An alternative interpretation, as a DB pulsator, does not seem viable for 
V803 Cen. Although there is one DB pulsator, PG1351+489 that has most of 
its power in one oscillation mode (at a period of 490 s) and appears almost 
mono-periodic (Winget, Nather \& Hill 1987), it and all other members of the 
class are in fact multi-periodic. The range of brightness of V803 Cen is over 
4 mag, showing that accretion luminosity dominates in the high state. That 
implies that any pulsations of the primary would be most visible in the low 
state; but the reverse is the case. Furthermore, the modulation is not always 
present at high state, which is a characteristic of DNOs, but not of pulsations 
of a degenerate star.

   We can demonstrate further aspects of the pulsations in V803 Cen that are 
in accord with DNO behaviour and not with DB pulsations. Table 1 lists 
observations that we have obtained in the high state; in addition to these we 
have drawn on archived observations published in O'Donoghue et al.~(1987) 
and O'Donoghue \& Kilkenny (1989). FTs and amplitude/O--C diagrams 
were computed for all of these runs. An example is given in Fig.~\ref{omc4038} and 
shows the occasional short-term instability of the pulsation, which is 
uncharacteristic of oscillations in degenerate pulsators.

     In Fig.~\ref{lc4035} we illustrate a light curve (run S4035 of O'Donoghue \& 
Kilkenny (1989)) of V803 Cen observed when the 176 s modulation was particularly 
prominent. In order to show this clearly we have prewhitened the light curve 
at the dominant 1611 s superhump modulation and its first three harmonics. The FT of the light curve
shown in Fig.~\ref{lc4035} is shown in Fig.~\ref{ft4035}. It shows 
how easy it is to detect the 176 s modulation when it is present at 
full amplitude. The first and second harmonics are very weakly present and 
appear in the mean profile of the modulation as a slight nonsinusoidality. 
Greater departure from sinusoidal shape is seen in some other runs, with 
higher harmonics present, but we have not found any light curves where the 
first harmonic alone is very prominent -- as would be the case for dominant 
two-pole accretion. Nor have we seen a subharmonic at $\sim$ 352 s, so we infer 
that the rotation period of the primary is $\sim$ 176 s.

\begin{figure}
\centerline{\hbox{\psfig{figure=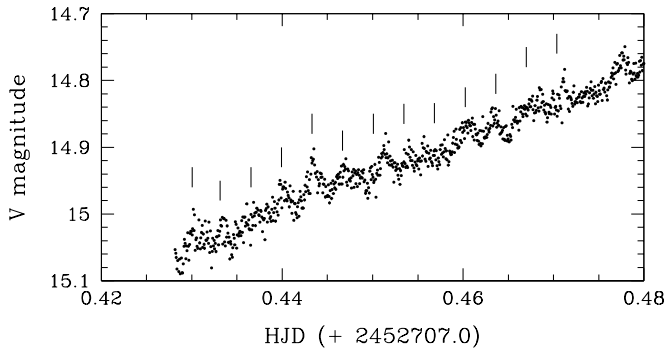,width=8.8cm}}}
  \caption{The light curve (the first $\sim$ 70 minutes of run S6852) of CR Boo.
The QPO maxima are marked.}
 \label{lc6852}
\end{figure}

\subsubsection{CR Boo}

\begin{figure}
\centerline{\hbox{\psfig{figure=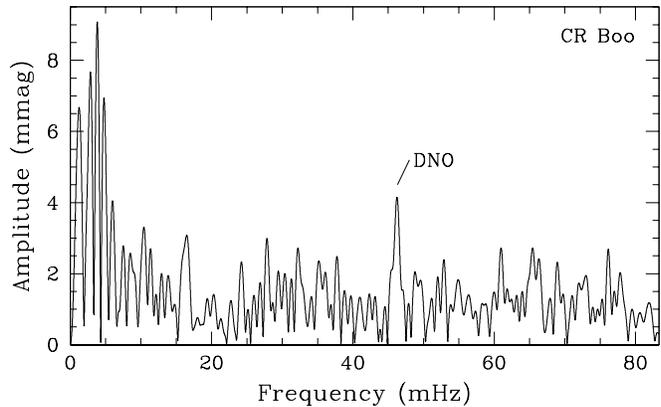,width=8.8cm}}}
  \caption{The Fourier Transform (data length $\sim$ 25 minutes) of a section of run S6852 of CR Boo.
The DNO frequency is marked.}
 \label{ftcrboo}
\end{figure}

The dominant modulation in CR Boo when in a high state is at 1490 s, with considerable
power in the 1st, 2nd and 3rd harmonics (Patterson et al.~1997). We have two long runs
on CR Boo (Table 1), the second (S6862) of which shows a light curve and FT similar 
to the common type, but the first (S6852) has both DNOs and QPOs, as seen in the light curve and FT 
(Figs.~\ref{lc6852} and \ref{ftcrboo}).

\begin{figure}
\centerline{\hbox{\psfig{figure=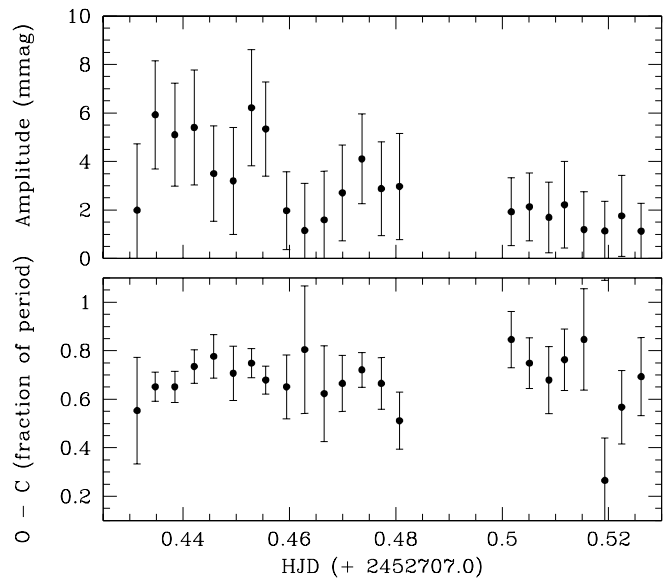,width=8.8cm}}}
  \caption{The O--C diagram (lower panel) and amplitude diagram (upper panel)
of the 61.9 s lpDNO seen in CR Boo.}
 \label{omc6852}
\end{figure}

The QPO signal is near, but not at the 4th harmonic. The latter whould be at 298 s, but
the peak in the FT is at 291 s, which is distinctly different. Furthermore, the apparent
QPO has large amplitude in the first half of the run, but disappears in the latter half,
which is unlike the behaviour of 4th harmonics seen in AM CVn stars (e.g., AM CVn itself:
Skillman et al.~1999).

\begin{figure*}
\centerline{\hbox{\psfig{figure=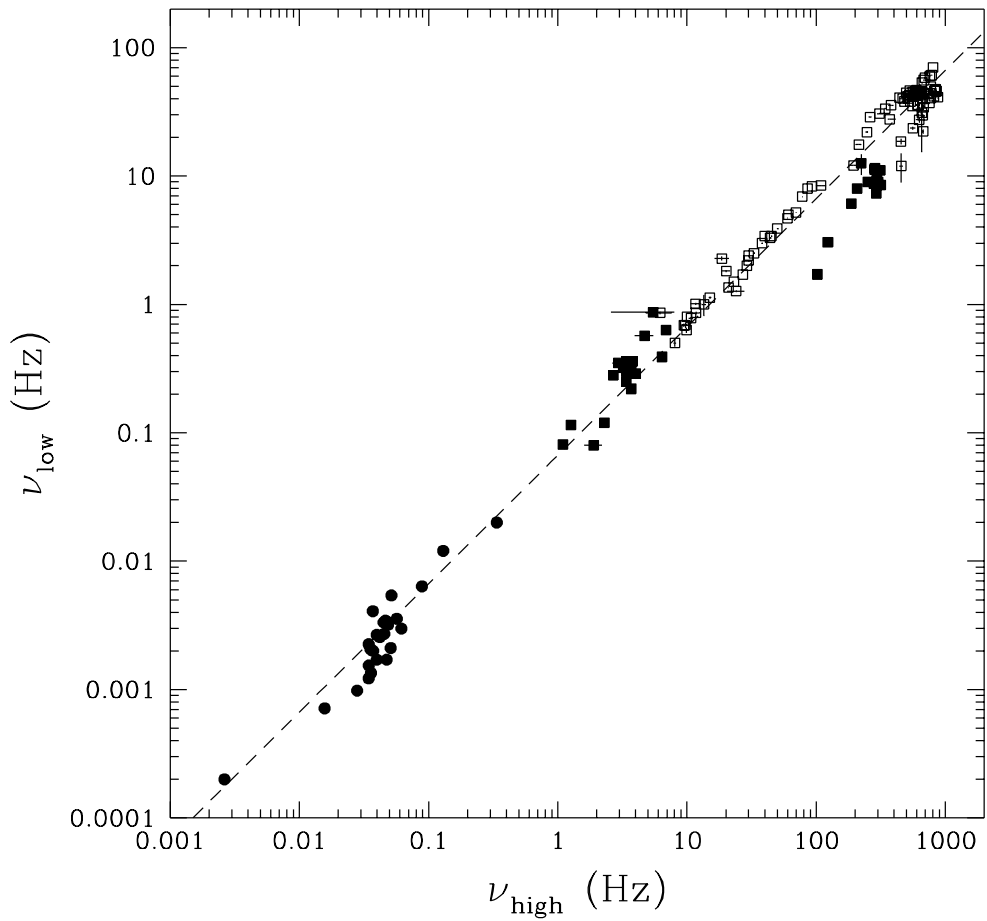,width=16.0cm}}}
  \caption{The Two-QPO diagram for X-Ray binaries (filled squares: black hole binaries;
open squares: neutron star binaries) and 26 CVs (filled circles). Each CV is only plotted once
in this diagram. The X-Ray binary data are from Belloni et al.~(2002) and were
kindly provided by T. Belloni. The dashed line marks $P_{QPO}/P_{DNO}$ = 15.}
 \label{twoqpo}
\end{figure*}

In the same run a DNO is present at 21.6 s, as seen in Fig.~\ref{ftcrboo}. This gives $R$ = 13.5.
The DNO seen in run S6852 is an independent confirmation of DNOs observed on three consecutive nights
at $\sim$ 22 s in CR Boo during the 1996 CBA campaign on CR Boo (Patterson, private communication).

In addition to the DNO and QPO observed in CR Boo (run S6852), there is
an lpDNO at 61.9~s in the same run. Fig.~\ref{omc6852} shows the phase/amplitude diagram
for the 61.9~s periodicity over the first $\sim$ 2.3 hours of run S6852.

\section{Further comparison between QPOs in CVs and X-Ray binaries}

      With our extended data set, obtained from observing the literature and 
the sky, we can generate a more detailed comparison between DNO and 
QPO periods in CVs. In Fig.~\ref{twoqpo} we show this on the same Two-QPO 
correlation diagram produced for X-Ray binaries (Belloni, Psaltis \& van der Klis 2002); the 
latter distinguishes between neutron star and black hole candidates. Some of 
the scatter in the CV part of the correlation may arise through ambiguities of 
period -- i.e. whether we are observing the primary periods or their first 
harmonics. Other scatter comes in a few cases from lack of simultaneity in 
observations of the DNOs and QPOs. However, the correlation is quite clear, 
and the whole relationship can be described by $P_{QPO} \sim 15 P_{DNO}$ both for CVs 
and X-Ray binaries, over a range of nearly six orders of magnitude in period or 
frequency. This correlation is not yet understood; Mauche (2002) has shown that
many popular models for QPOs in X-Ray binaries are excluded by this relationship.

    We are conscious of the possibility of bias in our selection of DNOs and 
QPOs that have been `accepted' into our list, but it was while attempting to 
avoid such a bias that we realised that there is strong evidence for an 
additional set of DNOs -- the lpDNOs described above. We currently see no 
clear evidence for maverick DNOs that fit neither of these categories.

    There is a possible additional similarity between CV and X-Ray binary 
oscillations. During X-Ray bursts in the latter, pulsations are seen which lie 
in the frequency range 300 -- 600 Hz (see Table 1 of van der Klis 2000), 
which increase in frequency slightly during the burst, and which are 
attributed to hot spots on the neutron star where gas expands and slips 
relative to the rotation of the underlying atmospheric layers. It is argued (van 
der Klis 2000) that at least in some cases there are two hot spots. If that were 
generally the case, then the true rotation frequencies would actually lie in the 
range 150 -- 300 Hz. The high frequency QPOs (plotted on the abscissa of 
Fig.~\ref{twoqpo}) in these same stars (Table 3 of van der Klis 2000) lie in the range 
500 -- 1100 Hz and are thus $\sim$ 4 times the supposed star rotation frequencies. 
This is the same relationship as seen in CVs, where the lpDNOs (interpreted 
above as white dwarf body rotation) have approximately four times the 
periods of the DNOs. The additional factor $\sim$ 4 between the periods of 
lpDNOs and QPOs in CVs is matched by the ratio of frequencies of rotation 
to low frequency QPOs in the X-Ray binaries.

   On the other hand, as can be seen in Table 2, there are examples in CVs 
where the first harmonic rather than the fundamental appears, and this can 
lead to $P_{QPO} \sim 2 P_{lpDNO}$, as seen in many of the X-Ray binary equivalent 
oscillations described above.

   The evolution of burst millisecond oscillations shows phase drifts and 
often sudden phase shifts that are similar to those of the lpDNOs (Muno, Chakrabaty \& Galloway 2002).

\section{Discussion and conclusion}

     In this paper we have extended the already rich phenomenology of CV 
DNOs and QPOs to include the recognition of a new type of DNO, the lpDNO. 
We have detected some double lpDNOs, whose splitting, as with the double DNOs,
suggests reprocessing off a travelling wave moving with the QPO period.
We have a possible first indication of such a sideband at the orbital frequency.
As the lpDNOs are a newly identified phenomenon we have taken care to 
illustrate them extensively in this paper. It is 
probable that there are many more aspects still to be uncovered.

    The differences in behaviour for outbursts of the same dwarf nova have 
yet to be understood. Why, during the December 1975 superoutburst of VW 
Hyi, the $\sim$ 88 s lpDNOs were so prominent for a week but only very 
occasionally and intermittently present in other outbursts, is a mystery. We 
note, however, that Hartmann et al.~(1999) have found that the X-Ray 
properties of VW Hyi also differ greatly from outburst to outburst. Both the 
X-Rays and the DNOs are associated with the inner disc and boundary layer, 
and both will be affected by the structure and strength of magnetic field 
connections from the disc to the primary and its equatorial belt, so this is 
where the differences are probably generated.

   Inter alia, we have shown that the helium-transferring systems in some 
cases have DNO and QPO properties similar to those of the hydrogen-rich 
CVs. This is first evidence, even if indirect, that the helium primaries of the 
AM CVn stars can be magnetic.

   An advantage that CV observations have over those of X-Ray binaries is 
that in CVs we often easily see the trains of DNO and QPO modulations, 
and can study the properties of individual cycles, whereas in the X-Ray 
binaries the low fluxes require statistical analyses to extract average 
properties. If the phenomena really do have the same origin, then further 
observations of CVs even with small and modest size telescopes are likely to 
assist understanding of the higher energy analogues. 

      At present the interpretation of the CV rapid oscillations in terms of 
magnetic accretion is speculative -- but we note that the behaviours are so 
complex that no simple explanations (e.g. white dwarf or accretion disc 
pulsations) are likely to be adequate. The existence of magnetically 
controlled accretion is in principle testable by the detection of polarisation 
modulated at the DNO period -- this might be accomplished for VW Hyi 
during the large amplitude DNOs at the end of outburst, which occur at 
roughly monthly intervals, but will probably require the use of a large 
telescope observing in the near infrared. Currently the situation is similar to 
that of the early development of the intermediate polar model (Patterson \& 
Price 1981; Warner, O'Donoghue \& Fairall 1981), which for a few years 
gave satisfactory explanations of the observed multiple photometric 
modulations but the magnetic nature was only verified by the discovery of 
circularly polarised I band flux in 1986 (Penning, Schmidt \& Liebert 1986).

     Van der Klis (2000), in connection with the two-QPO correlation in X-Ray 
binaries and the concomitant variations of the frequencies, remarked 
that they ``essentially imply that the phenomena are generated in the 
accretion disc around any low-magnetic field compact object''. Our 
observations of similar phenomena in (presumably) low-field CVs appear to 
make this conclusion even more widely applicable.

\section*{Acknowledgments}

         The University of Cape Town funds BW's research; PAW is supported 
by strategic funds made available to BW by the University and by research 
funds from the National Research Foundation. MLP acknowledges financial support
from the Department of Labour's National Skills Fund.  We 
thank Dave Kilkenny for alerting us to the potential interest of EC 2117-54. 
T. Belloni kindly sent us the data on X-Ray binaries that appear in Fig.~\ref{twoqpo}.

\end{document}